\documentclass[twocolumn,aps,prb,superscriptaddress,showpacs,floatfix]{revtex4-1}
\usepackage{hyperref}
\usepackage{graphicx}
\usepackage{tikz}
\usepackage{float}
\usepackage{amsmath}
\usepackage{amssymb}
\usepackage{bm}
\usepackage[normalem]{ulem}

\newcommand{\mb}[1]{ { \mbox{\boldmath{$#1$}}}  } 
 
\begin{document}
\title{Interplay between electron pairing and Dicke effect
       in triple quantum dot structures}

\author{S. G\l{}odzik}
\affiliation{Institute of Physics, M.\ Curie Sk\l odowska University, 20-031 Lublin, Poland}

\author{K.P. W\'ojcik}
\affiliation{Faculty of Physics, A.\ Mickiewicz University, 61-614 Pozna\'n, Poland}

\author{I. Weymann}
\email{weymann@amu.edu.pl}
\affiliation{Faculty of Physics, A.\ Mickiewicz University, 61-614 Pozna\'n, Poland}

\author{T. Doma\'nski}
\email{doman@kft.umcs.lublin.pl}
\affiliation{Institute of Physics, M.\ Curie Sk\l odowska University, 20-031 Lublin, Poland}

\date{\today}

\begin{abstract}
We study the influence of the proximity-induced pairing on electronic version of the Dicke 
effect in a heterostructure, comprising three quantum dots vertically coupled between 
the metallic and superconducting leads. We discuss a feasible experimental procedure
for detecting the narrow/broad (subradiant/superradiant) contributions by means of 
the subgap Andreev spectroscopy. In the Kondo regime and for small energy level 
detuning the Dicke effect is manifested in the differential conductance.
\end{abstract}  



\maketitle

\section{Introduction}

Triple quantum dots coupled to the reservoirs of mobile electrons enable 
realization of the electronic Dicke effect \cite{Orellana2006}. The original 
phenomenon, known in quantum optics, manifests itself by the narrow 
({\em subradiant}) and broad ({\em superradiant}) lineshapes spontaneously 
emitted by atoms linked on a distance smaller than a characteristic 
wavelength \cite{Dicke1953}. Early prototypes of its electronic 
counterpart have been considered by several groups 
\cite{Raikh1994,Wunsch2003,Vorrath2003,Orellana2004,Brandes2005}.

\begin{figure}
\includegraphics[width=0.9\linewidth]{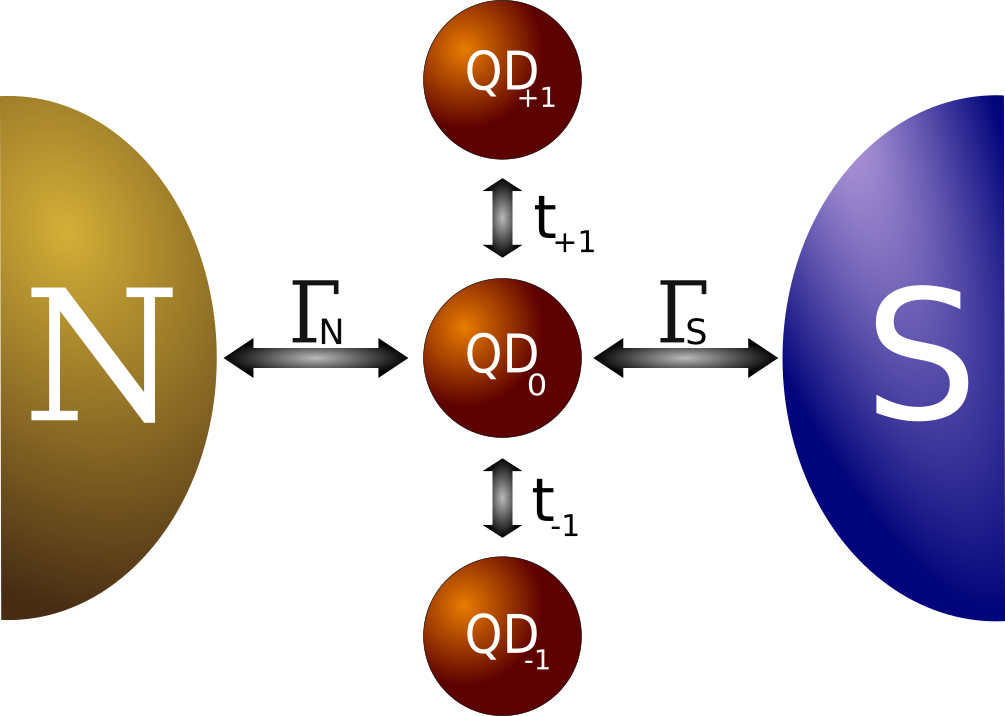}
\caption{Schematic view of three quantum dots (QD$_{j}$) arranged vertically  
between the normal (N) and superconducting (S) electrodes. The central 
quantum dot, QD$_{0}$, is coupled by $\Gamma_{\beta}$ to the external reservoirs and   
by $t_{\pm 1}$ to the side-attached quantum dots, QD$_{\pm 1}$.}
\label{schematic_view}
\end{figure}

In nanostructures, where the central quantum dot (QD$_{0}$) with two 
side-attached dots (QD$_{\pm 1}$) are arranged in a crossed bar configuration 
(Fig.\ \ref{schematic_view}), the sub- and superradiant 
contributions can be achieved either upon increasing the inter-dot coupling 
$t_{\pm 1}$ or via tuning the quantum dot energy levels $\epsilon_{\pm 1} 
\!\rightarrow \! \epsilon_{0}$. Such scenario has been investigated 
for heterojunctions with both normal (conducting) electrodes 
\cite{guevara2006,trocha2008a,trocha2008b,vernek2010,baruselli2013,wang2013}. 
In particular, an interplay between the Kondo and Dicke effects, 
manifested in the differential conductance, has been addressed
\cite{trocha2008b,vernek2010}. Moreover, it has been shown that the electronic 
Dicke effect substantially enhances the thermoelectric properties 
and can violate the Wiedemann-Franz law \cite{wang2013}. 

Selected aspects of the electronic Dicke effect have been confronted also with 
superconductivity, considering the Andreev \cite{bai2010a,bai2010,ye2015,xu2016} 
and Josephson-type \cite{yi2016,wang2016} spectroscopies. To the best of  our 
knowledge, however, a thorough description of the relationship between 
the induced electron pairing, the Dicke effect and the strong correlations is 
missing. We address this problem here, focusing on the low-energy $|\omega| 
<\Delta$ (subgap) regime of the Andreev-type setup. Our main purpose is to 
establish knowledge on how the electron pairing and correlation effects are affected by 
the side-attached quantum dots, QD$_{\pm 1}$, ranging from the interferometric
(weak $t_{\pm 1}$ coupling)
to the molecular (strong inter-dot coupling) limits. Our studies 
reveal strong redistribution of the spectral weights (although  
manifested  differently for these extremes),
suppressing the low-energy (subradiant) states.
Transfer of this spectral weight has an influence 
on the subgap Kondo effect, which can be observed experimentally
by the zero-bias Andreev conductance.

The paper is organized as follows. In Sec.\ \ref{sec:II} we introduce 
the microscopic model and describe the method accounting for the induced 
electron pairing. Sec.\ \ref{sec:III} corresponds to the case of uncorrelated 
quantum dots in the deep subgap regime, studying evolution of the central 
quantum dot spectrum from the weak to strong interdot coupling. Next, 
in Sec.\ \ref{sec:IV}, we discuss the correlation effects in the subgap 
Kondo regime.  Finally we summarize the results and present 
the conclusions.

\section{Microscopic model\label{sec:II}}

The central quantum dot, QD$_{0}$, placed between the normal (N) and 
superconducting (S) electrodes and side-attached to the quantum dots, 
QD$_{\pm 1}$, as shown in Fig.\ \ref{schematic_view}, can be modeled 
by the Anderson-type Hamiltonian
\begin{equation}
\hat{H}=\hat{H}_{QD}+\hat{H}_N+\hat{H}_{S}+\hat{H}_{QD-N} +\hat{H}_{QD-S}.
\label{model}
\end{equation}
The set of three quantum dots can be described by
\begin{eqnarray}
\hat{H}_{QD} & = & \sum_{\sigma,j} \epsilon_{j} \hat{d}^\dagger_{j\sigma}
\hat{d}_{j\sigma} + \sum_{\sigma,{j=\pm 1}} \left( t_{j} \hat{d}^\dagger_{0\sigma} 
\hat{d}_{j\sigma} + \mbox{\rm h.c.} \right) \nonumber \\ 
&+& \sum_{j} U_{j} \hat{n}_{j\uparrow} \hat{n}_{j\downarrow} ,
\label{three_dots}
\end{eqnarray}
where $\hat{d}_{j\sigma}^{(\dagger)}$ annihilates (creates) electron of 
$j$-th quantum dot with energy $\epsilon_{j}$ and spin $\sigma=\uparrow,
\downarrow$. Hybridization between the quantum dots is characterized
by the hopping integral $t_{\pm 1}$. We denote the number operator by 
$\hat{n}_{j\sigma}= \hat{d}_{j\sigma}^{\dagger}\hat{d}_{j\sigma}$ 
and $U_{j}$ stands for the Coulomb potential which is responsible 
for correlation effects. 

We treat the normal (metallic) lead electrons as a free fermion gas $\hat{H}_N
=\sum_{{\bf k},\sigma} \xi_{kN} \hat{c}^\dagger_{{\bf k}N\sigma}
\hat{c}_{{\bf k}N\sigma}$ and describe the superconductor by 
the BCS model $\hat{H}_{S} = \sum_{{\bf k},\sigma} \xi_{{\bf k}S} 
\hat{c}^\dagger_{{\bf k}S\sigma} \hat{c}_{{\bf k}S\sigma} - 
\sum_{\bf k} \Delta \left( \hat{c}^\dagger_{{\bf k}S\uparrow}
\hat{c}^\dagger_{-kS\downarrow} +{\mbox{\rm h.c.}} \right)$ with
the isotropic energy gap $\Delta$.
Operators $\hat{c}^{(\dagger)}_{{\bf k}\beta\sigma}$ refer to the mobile 
electrons of external ($\beta\!=\!N,S$) electrodes whose energies 
$\xi_{{\bf k}\beta}=\epsilon_{\bf k} -\mu_{\beta}$ are expressed with 
respect to the chemical potentials $\mu_{\beta}$. For convenience we 
choose $\mu_{S}=0$ as a reference level. Tunneling between the central 
dot and the external leads is described by $\hat{H}_{QD-\beta} = \sum_{{\bf k},\sigma} 
\left( V_{{\bf k} \beta}\hat{c}^\dagger_{{\bf k} \beta \sigma}\hat{d}_{0\sigma}  
+ {\mbox{\rm h.c.}} \right)$, where $V_{{\bf k} \beta}$ denote the matrix elements. 
Focusing on the subgap quasiparticle states we apply the wide-band limit 
approximation, assuming the energy independent couplings $\Gamma_\beta=
2\pi \sum_{\bf k} \left| V_{{\bf k}\beta} \right|^{2} \delta(\omega-
\epsilon_{{\bf k}\beta})$. 

\subsection{Superconducting proximity effect}

Measurable properties of our heterostructure predominantly depend on the
effective spectrum of the central quantum dot, which results from: (i) the proximity 
induced pairing, (ii) electron correlations and (iii) influence of the side-attached 
quantum dots QD$_{\pm 1}$. The superconducting proximity effect mixes
the particle and hole degrees of freedom, therefore we have to 
introduce the matrix Green's function
\begin{equation}
{\mb G}_{j}(t,t')= 
\begin{pmatrix}
\langle\langle \hat{d}_{j\uparrow}(t) ; \hat{d}_{j\uparrow}^{\dagger}(t')\rangle\rangle & 
\langle\langle \hat{d}_{j\uparrow}(t) ; \hat{d}_{j\downarrow}(t')\rangle\rangle \\
 \langle\langle \hat{d}_{j\downarrow}^\dagger(t) ; \hat{d}_{j\uparrow}^\dagger(t')\rangle\rangle & 
\langle\langle \hat{d}_{j\downarrow}^\dagger(t) ; \hat{d}_{j\downarrow}(t')\rangle\rangle
\end{pmatrix} ,
\label{GF_definition}
\end{equation}
where $\langle\langle \hat{A}(t);\hat{B}(t')\rangle\rangle = -i\Theta(t\!-\!t')
\langle \left[ \hat{A}(t), \hat{B}(t')\right] \rangle$ is the retarded fermion 
propagator. In stationary case (for time-independent Hamiltonian) the Green's 
function (\ref{GF_definition}) depends on $t\!-\!t'\equiv\tau$ and its Fourier 
transform ${\mb G}_{j}(\omega) \equiv \int d\tau  e^{-i\omega \tau} {\mb G}_{j}
(\tau)$ obeys the Dyson equation
\begin{equation}
\left[ {\mb G}_{j}(\omega) \right]^{-1}=
\begin{pmatrix}
\omega-\epsilon_j & 0 \\
0 & \omega+\epsilon_j
\end{pmatrix}
-{\mb \Sigma}_{j}(\omega) .
\label{Dyson}
\end{equation}
The selfenergy matrix ${\mb \Sigma}_{j}(\omega)$ describes influence of 
the inter-dot couplings, the external leads, and the correlations. In 
general, its analytic form is unknown.

\begin{figure}
\includegraphics[width=0.9\linewidth]{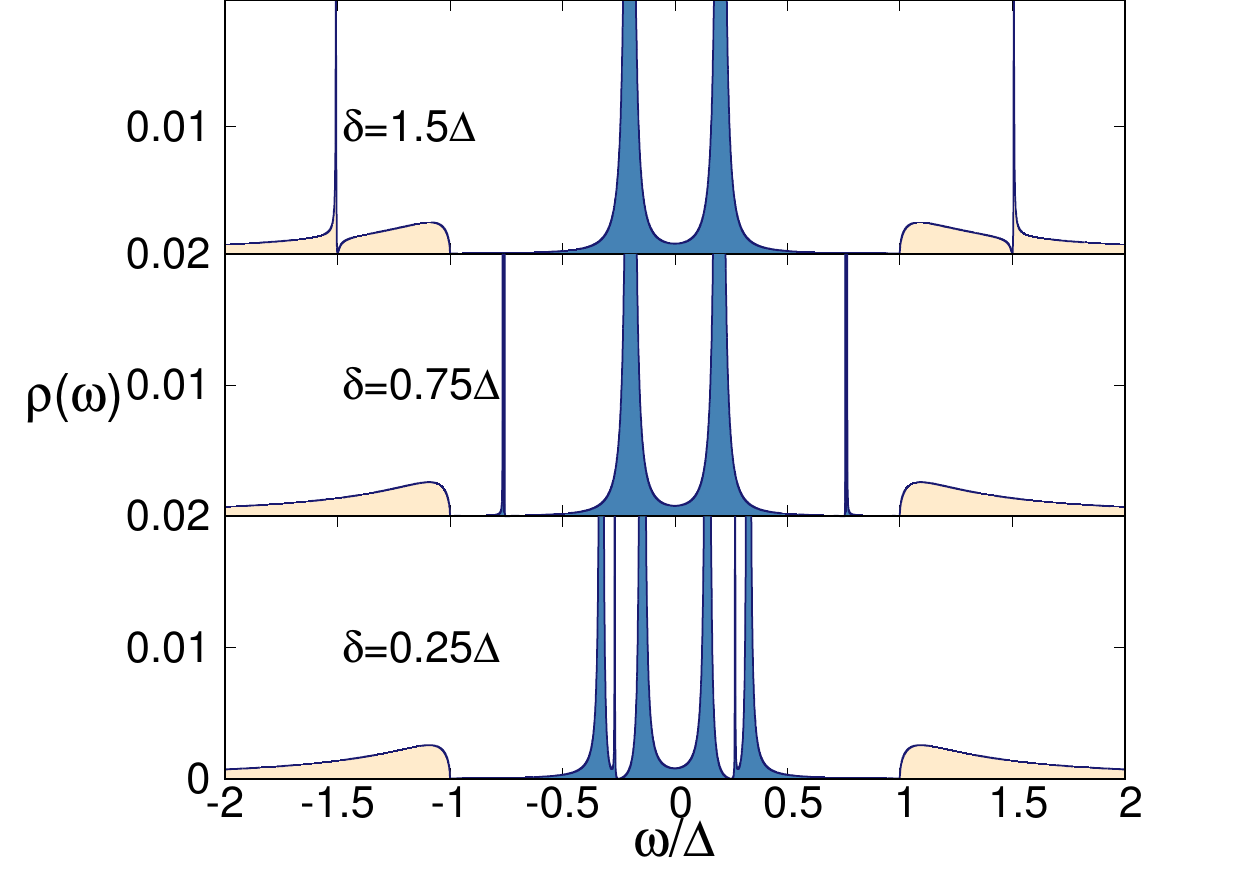}
\caption{Spectral function $\rho(\omega)$ [in units of $\frac{2}{\pi\Gamma_{N}}$] 
of the central dot obtained for $\epsilon_0=0$, $\Gamma_{S}/\Delta=0.5$, 
$\Gamma_{N}/\Delta=0.01$, $U_{0}=0$, $t/\Delta=0.15$ and representative detunings: 
$\delta/\Delta=1.5$ (top), $0.75$ (middle), $0.25$ (bottom panel).}
\label{finite_Delta}
\end{figure}

\subsection{Features of a weak inter-dot coupling}

It is instructive to analyze first how the side-attached quantum dots 
come along with the proximity induced electron pairing, neglecting 
the correlations $U_{j}=0$. The selfenergy of uncorrelated QD$_{0}$ 
is given by
%
\begin{eqnarray}
{\mb \Sigma}^{U=0}_0(\omega) &=& \begin{pmatrix}
\frac{-i\Gamma_N}{2}-\frac{i\Gamma_S}{2}\tilde{\rho}(\omega) 
& - \frac{i\Gamma_S}{2}\tilde{\rho}(\omega)\frac{\Delta}{\omega}  \\  
-\frac{i\Gamma_S}{2}\tilde{\rho}(\omega)\frac{\Delta}{\omega} 
& \frac{-i\Gamma_N}{2}-\frac{i\Gamma_S}{2}\tilde{\rho}(\omega)  
\end{pmatrix}
\nonumber \\
&+& \sum_{j=\pm 1} \begin{pmatrix}
\frac{t_{j}^2}{\omega-\epsilon_{j}} & 0 \\
0 & \frac{t_{j}^2}{\omega+\epsilon_{j}} 
\end{pmatrix} ,
\label{selfenergy_free} 
\end{eqnarray}
%
where  
 \begin{equation}
 \tilde{\rho}(\omega)= \left\{
\begin{array}{ll} 
\frac{\omega}{\sqrt{\Delta^{2}-\omega^{2}}}
& \mbox{\rm for }  |\omega| \leq \Delta , \\
\frac{i\;|\omega|}{\sqrt{\omega^{2}-\Delta^{2}}}
& \mbox{\rm for }  |\omega| > \Delta .
\end{array} \right.
\end{equation}

Let us inspect the spectral function of QD$_{0}$ 
\begin{equation}
\rho(\omega)=-\frac{1}{\pi} \; \mbox{\rm Im} \left\{ G_{0,11}
(\omega+i0^{+}) \right\} ,
\end{equation}
assuming  the side-attached quantum dots to be weakly coupled to the central 
dot.  Following the previous studies of three quantum dots on interface 
between two metallic electrodes \cite{trocha2008a,trocha2008b,vernek2010} 
we impose $t_{-1}=t_{+1}\equiv t$ and define the energy detuning 
$\epsilon_{+1}-\epsilon_{0}=\epsilon_{0}-\epsilon_{-1}\equiv \delta$. 
Figure \ref{finite_Delta} shows $\rho(\omega)$ for the asymmetric couplings 
$\Gamma_{S}>\Gamma_{N}$, when the quasiparticle states of the subgap regime 
(marked by blue color in Fig.\ \ref{finite_Delta})
are sufficiently narrow (long-lived). For the large 
detuning $\delta > \Delta$ (top panel) we observe two Fano-type resonances 
appearing outside the superconducting gap at $\omega=\pm\delta$. 
For the moderate detuning $\delta =0.75 \Delta$ (middle panel) there 
appear some features inside the superconducting gap, but they no longer resemble 
Fano-type lineshapes. For the very small detuning $\delta=0.25\Delta$ 
(bottom panel), a rather complicated subgap structure emerges. 
To clarify its physical origin, we explore in section \ref{sec:III} 
the deep subgap regime $|\omega|\ll\Delta$.

\section{Subgap Dicke effect vs pairing\label{sec:III}}

In this part we study in more detail the extreme subgap region $|\omega|\ll\Delta$, 
for which the selfenergy (\ref{selfenergy_free}) simplifies to
\begin{equation}
\lim_{|\omega|\ll\Delta}{\mb \Sigma}^{U\!=\!0}_0(\omega)=\begin{pmatrix}
\frac{-i\Gamma_N}{2} +\sum_{j}\frac{t_{j}^2}{\omega-\epsilon_{j}}
& -\frac{\Gamma_S}{2}  \\  
-\frac{\Gamma_S}{2}& \frac{-i\Gamma_N}{2} 
+\sum_{j}\frac{t_{j}^2}{\omega+\epsilon_{j}}
\end{pmatrix} 
\label{proximized_QD0}
\end{equation}
with summation running over $j = \pm 1$. 
The presence of the superconducting reservoir shows up in the selfenergy 
(\ref{proximized_QD0}) through the static off-diagonal terms, which 
can be interpreted as the induced on-dot pairing potential \cite{Bauer-2007}. 

\subsection{From interferometric to molecular regions}

\begin{figure}
\includegraphics[width=0.92\linewidth]{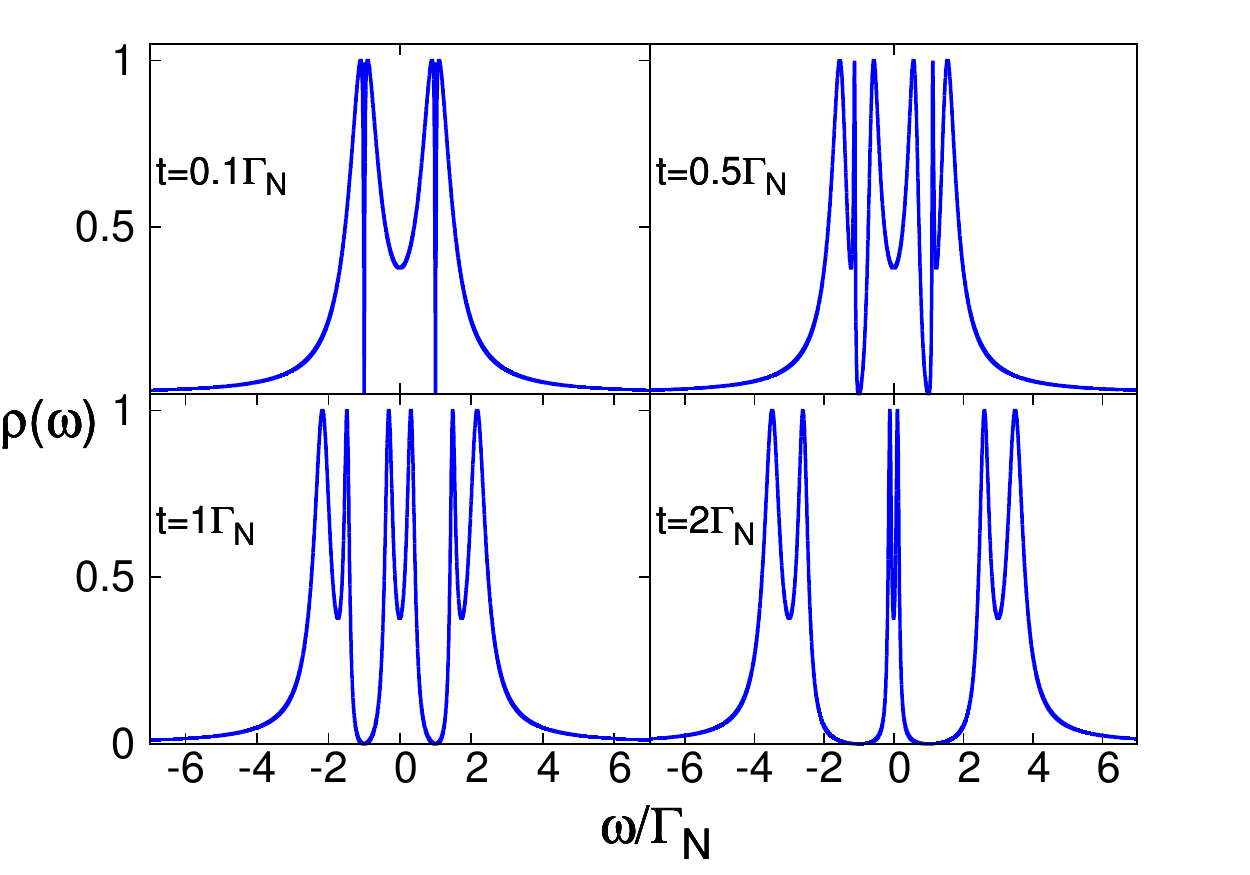}
\caption{Electronic spectrum of QD$_{0}$ obtained for $\Gamma_S=2\Gamma_N$, 
$\delta=\Gamma_{N}$, $U_{j}=0$, $\varepsilon_{0}=0$ and various interdot couplings $t$.}
\label{spectrum_QD_central}
\end{figure}

\begin{figure}
\includegraphics[width=0.88\linewidth]{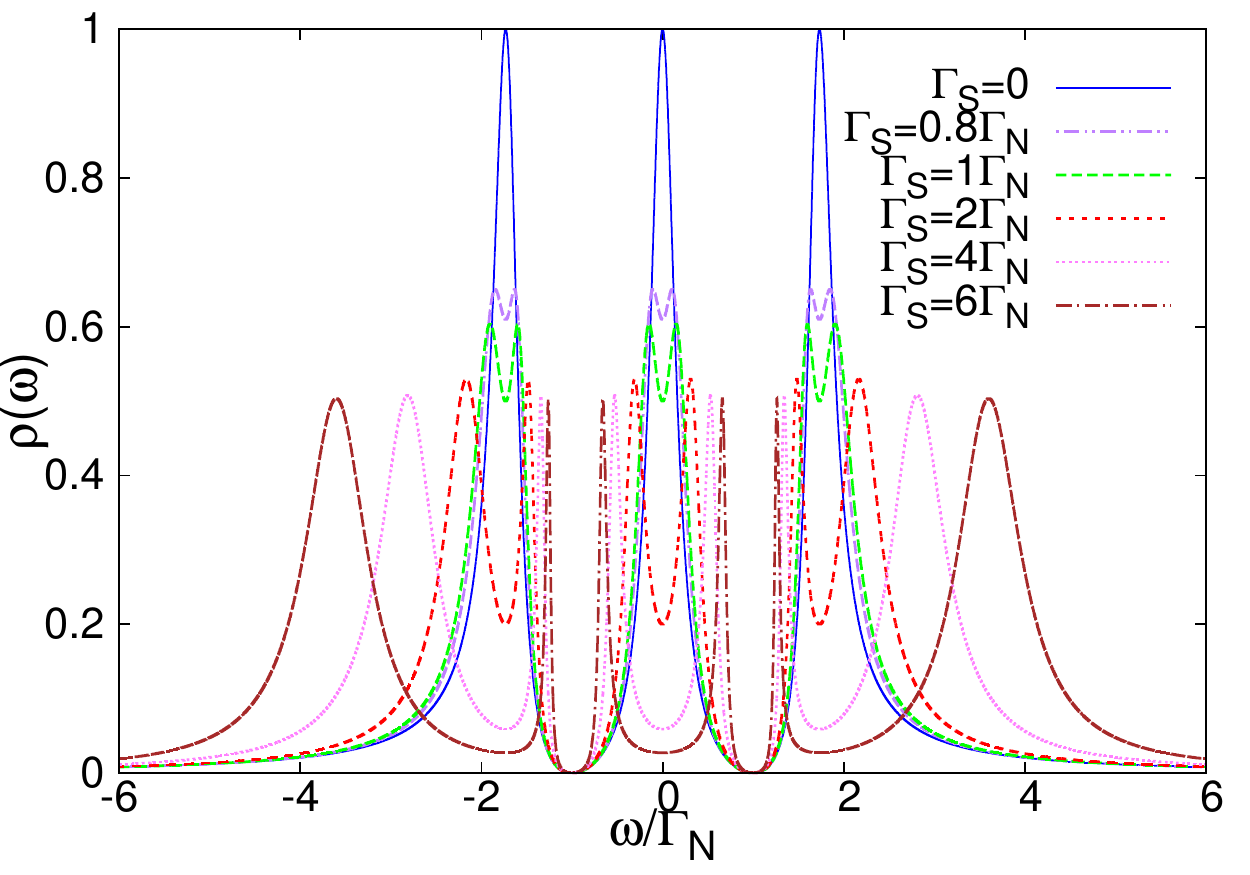}
\caption{Spectral function $\rho(\omega)$ of the uncorrelated QD$_{0}$ 
obtained in the molecular region $t=\Gamma_N$ for $\delta=\Gamma_{N}$, 
$\varepsilon_{0}=0$, $U_{j}=0$ and various couplings $\Gamma_S$, as indicated.}
\label{figure_splitting}
\end{figure}

\begin{figure}
\includegraphics[width=0.85\linewidth]{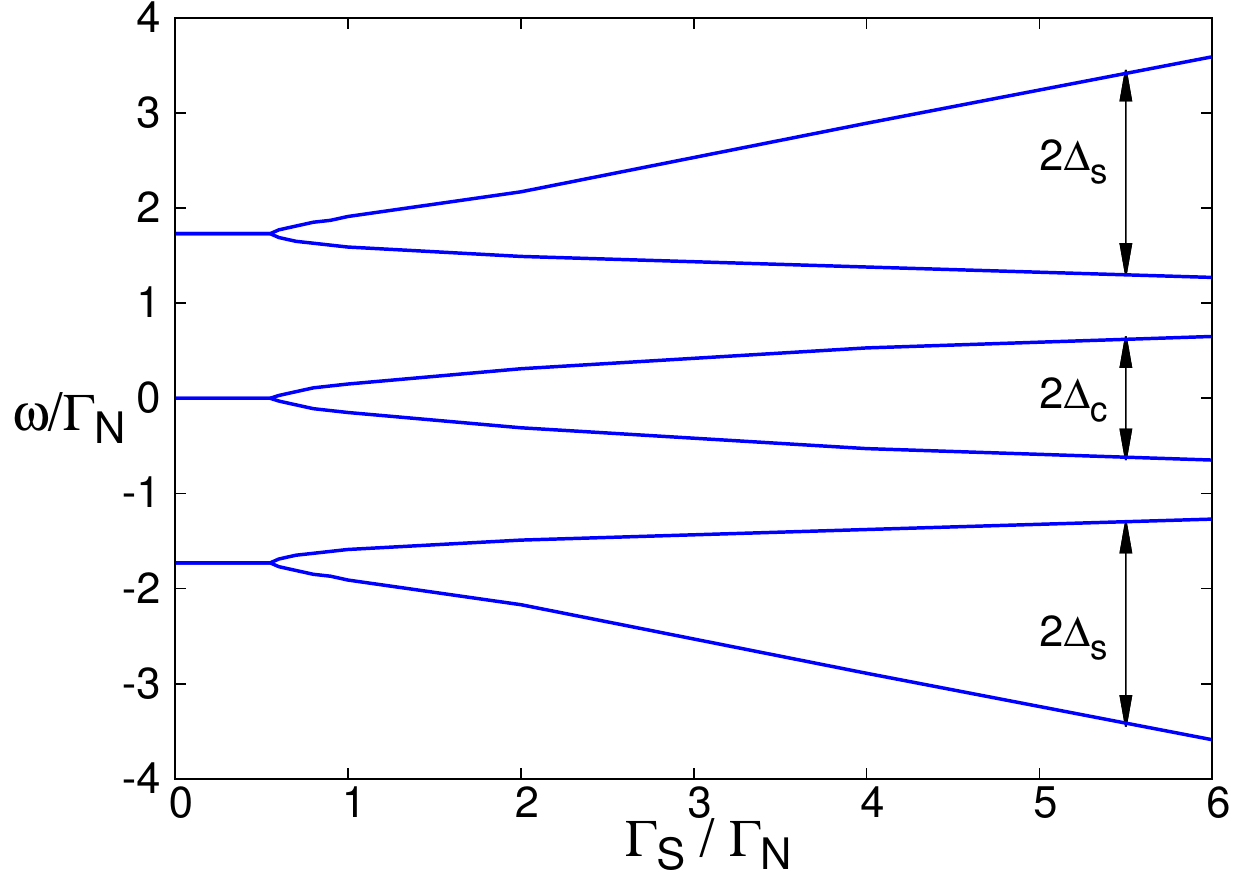}
\caption{Splitting of the subradiant ($\Delta_{c}$) and superradiant 
($\Delta_{s}$) quasiparticle states caused by the superconducting proximity 
effect for $t=\Gamma_N$, $\delta=\Gamma_{N}$, $\varepsilon_{0}=0$, $U_{j}=0$.}
\label{induced_gap}
\end{figure}

Figure \ref{spectrum_QD_central} presents the spectral function $\rho(\omega)$ 
obtained for  $\Gamma_{S}=2\Gamma_{N}$ and several values of the interdot 
coupling, ranging from the interferometric 
(small $t$) to the molecular (large $t$) regimes. For the weak  coupling $t=0.1
\Gamma_{N}$ we observe the Fano-type lineshapes at $\epsilon_{\pm 1}$ appearing
on top of the Andreev quasiparticles that are centered at $\pm \sqrt{\epsilon_{0}^{2}
+(\Gamma_{S}/2)^{2}}$. With increasing $t$ the spectrum gradually evolves to 
the `molecular' structure, characterized by the subradiant (narrow central) 
quasiparticle and superradiant (broad side-peaks) states. Similar tendency 
has been reported for the heterojunction with both normal leads 
\cite{trocha2008a,trocha2008b,vernek2010}. In the present case, however, we
observe additional qualitative changes caused by the proximity effect. Figure
\ref{figure_splitting} shows the evolution of the spectral function $\rho(\omega)$ 
with respect to $\Gamma_S$. At some critical coupling $\Gamma_{S}\approx 0.6\Gamma_{N}$ 
the sub- and superradiant states effectively split due to the on-dot electron pairing. 
We denote these splittings by $\Delta_{c}$ for the central peak and by $\Delta_{s}$ 
for the side peaks, respectively. Their magnitudes are displayed in Fig.\ 
\ref{induced_gap}.

\begin{figure}
\includegraphics[width=0.92\linewidth]{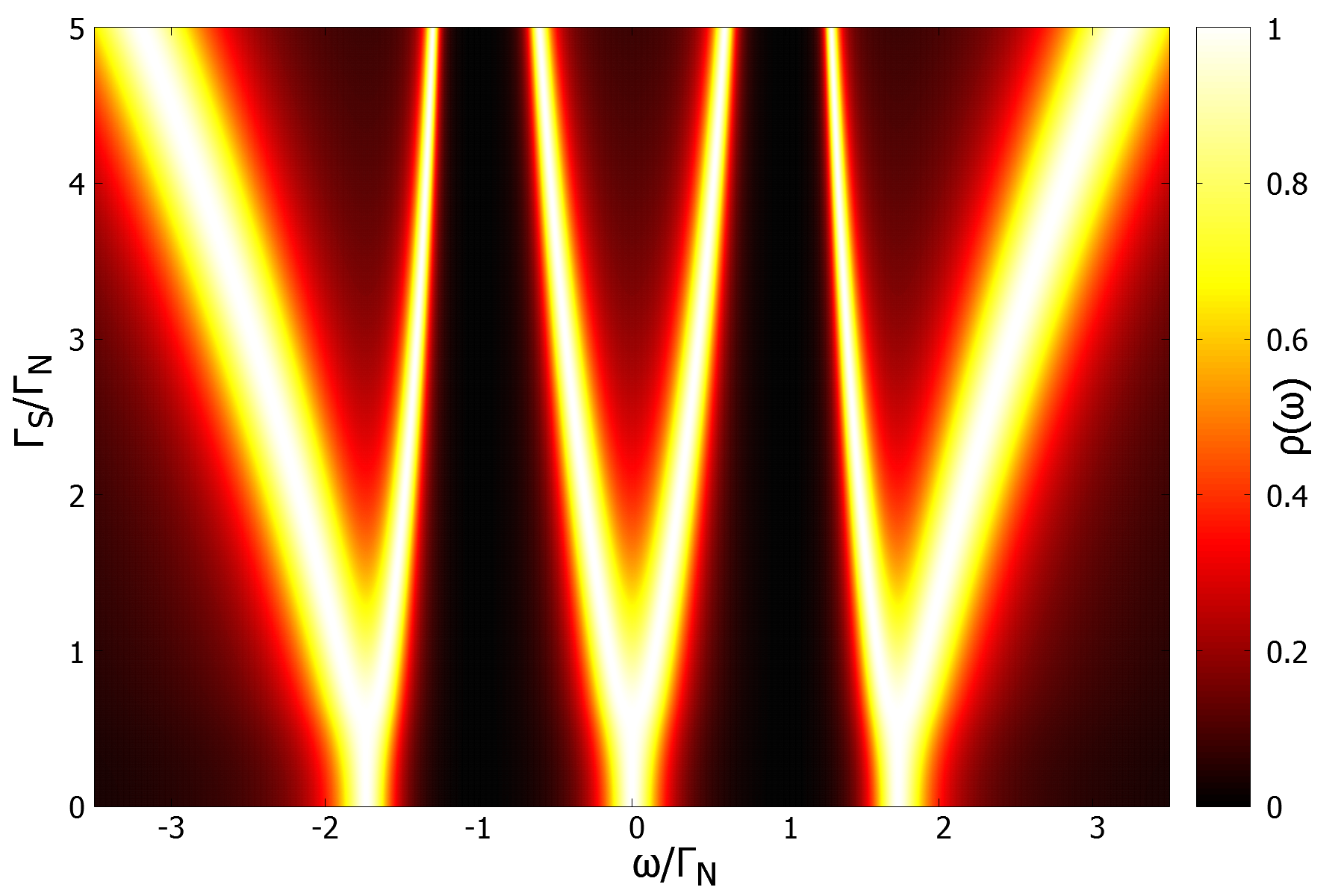}
\includegraphics[width=0.92\linewidth]{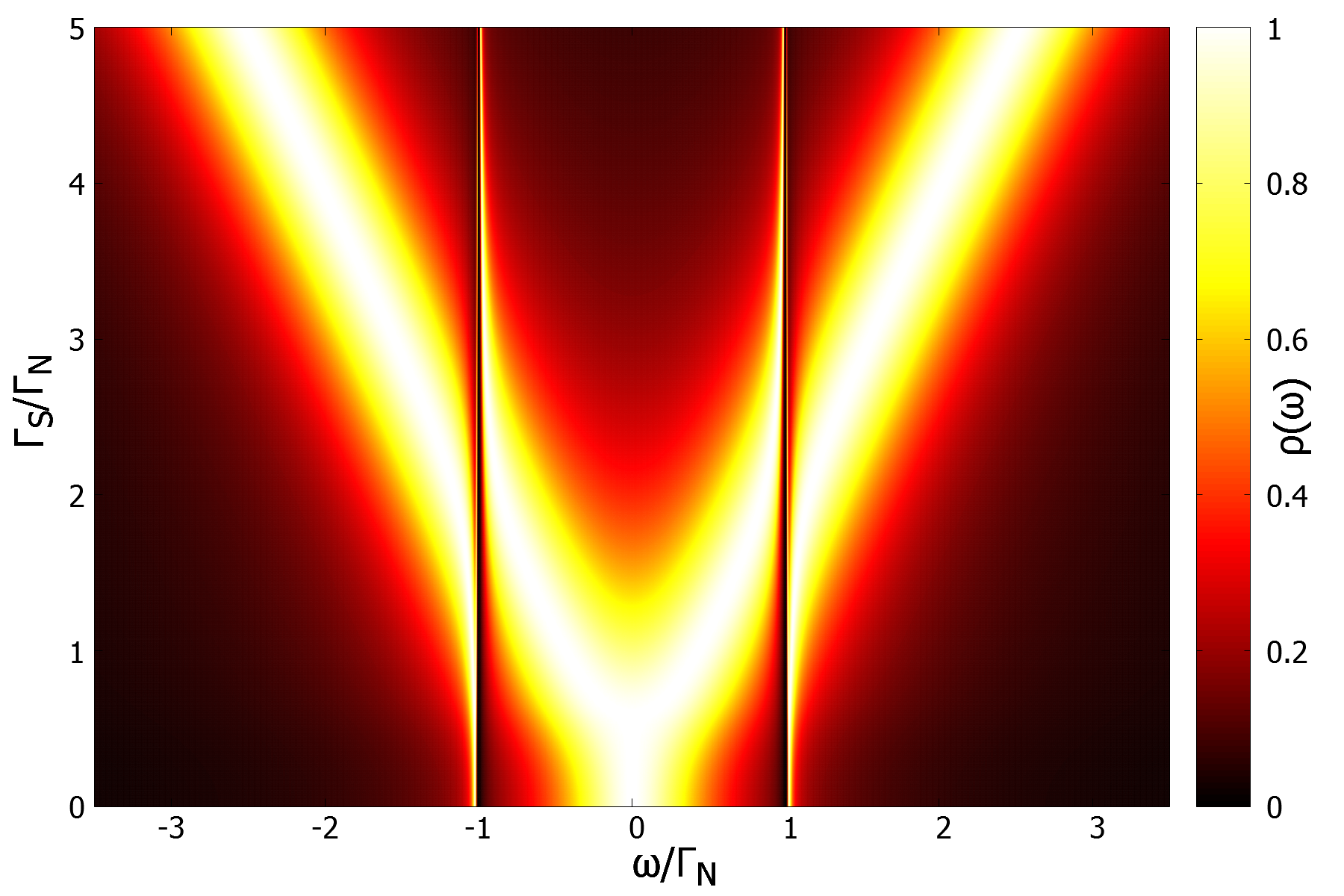}
\caption{Electronic spectrum of the central QD$_{0}$  in the molecular 
$t=1\Gamma_{N}$ (upper panel) and interferometric $t=0.15\Gamma_{N}$ 
(bottom panel) regions obtained for $U_j=0$, $\varepsilon_{0}=0$, 
$\delta=\Gamma_{N}$.}
\label{figure_new}
\end{figure}

We notice, that particle-hole splitting of the central (subradiant) 
peak differs from the corresponding effect in the side (superradiant) peaks,
see the upper panel of Fig.\ \ref{figure_new}
which shows the spectral function of the middle quantum dot QD$_0$.
Symmetric shape of the 
subradiant quasiparticle is perfectly preserved, but with increasing 
$\Gamma_{S}$ its internal splitting is bounded from above ($\Delta_{c}
\rightarrow \delta$). Such limitation comes from the destructive quantum 
interference, which depletes the electronic states around $\epsilon_{\pm 1}$. 
On the other hand, the superradiant quasiparticle peaks are not much 
affected by any constraints, therefore $\Delta_{s}$ monotonously grows 
with increasing $\Gamma_{S}$. We observe, however, that such superradiant 
states acquire asymmetric shape with the narrow structure slightly 
outside $|\epsilon_{\pm 1}|$  and another broader peak in the high energy 
regime. In the extremely strong  $\Gamma_{S}$ coupling limit, the high energy 
peaks absorb majority of the spectral weight.

\begin{figure}
\includegraphics[width=0.95\linewidth]{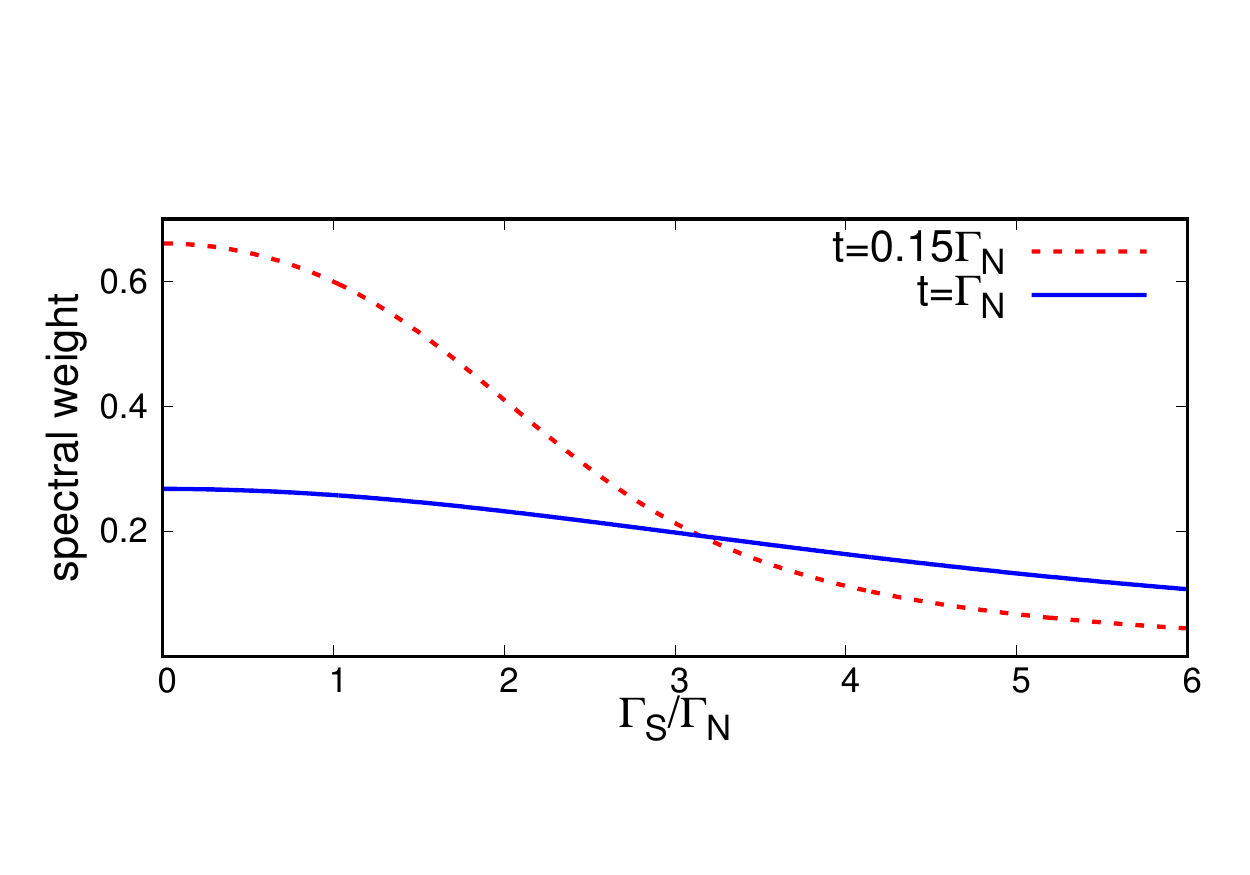}
\caption{Spectral weight of the low energy electronic states for
$\omega \in \left[ \epsilon_{-1},\epsilon_{+1} \right]$
caused by the  electron pairing for the interferometric 
(dashed line) and molecular (solid line) regions.}
\label{transfer_of_spectral_weight}
\end{figure}

On the other hand, in the interferometric regime (see bottom panel 
in Fig.\ \ref{figure_new}) we observe the Andreev quasiparticle
states (centered around $\pm \Gamma_{S}/2$ and their broadening equal
$\Gamma_{N}$) with the Fano-type lineshapes appearing
at $\omega=\epsilon_{\pm 1}$. Total spectral weight contained 
in the regime $\omega \in \left[ \epsilon_{-1},\epsilon_{+1} 
\right]$ is gradually washed out with increasing $\Gamma_{S}$. Such 
transfer of the spectral weight for the molecular and interferometric 
cases is displayed in Fig.\ \ref{transfer_of_spectral_weight}. In 
both cases the induced electron pairing depletes the low-energy  
quasiparticle states by transferring their spectral weight towards 
the higher energy quasiparticle states. Section \ref{sec:IV} 
shows that this process constructively affects the Kondo effect.

\subsection{Subgap tunneling conductance}

Any experimental verification of the subgap energy spectrum can be performed by 
measuring the tunneling current, induced under nonequilibrium conditions
$\mu_{N}-\mu_{S}=eV$ (where $V$ is an applied voltage). At low voltage
 the subgap current is provided solely by the anomalous Andreev channel,  
when electrons are scattered back to $N$ electrode as holes, injecting 
the Copper pairs to superconducting electrode. Within the Landauer 
approach such current can be expressed by
\begin{eqnarray}
I(V) = \frac{2e}{h} \int d\omega \; T_A(\omega) 
\left[ f_{FD}(\omega-eV) - f_{FD}(\omega+eV) \right] , 
\nonumber
\end{eqnarray}
where $f_{FD}(\omega) = \left[ 1 + 
\mbox{\rm exp}\left( \omega/k_{B}T \right) \right]^{-1}$ is the 
Fermi-Dirac distribution function. The Andreev transmittance 
$T_A(\omega)=\Gamma_N^2 \left| {\mb G}_{0,12}(\omega) \right|^{2}$ is
a quantitative measure of the proximity induced pairing which indirectly 
probes the subgap electronic spectrum, although in a symmetrized 
manner, because the particle and hole degrees of freedom equally 
contribute to such transport channel. 

\begin{figure}
\includegraphics[width=0.92\linewidth]{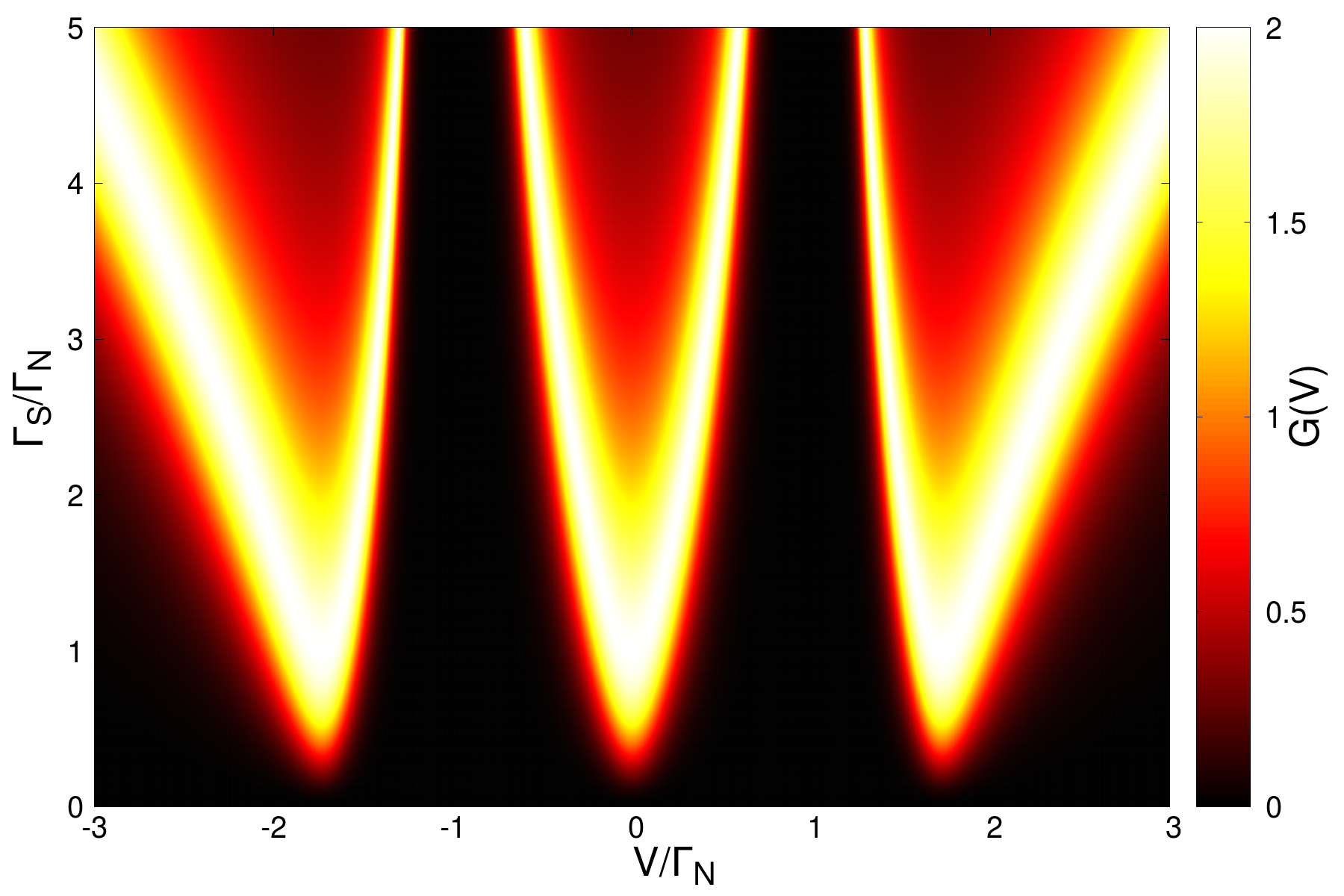}
\includegraphics[width=0.92\linewidth]{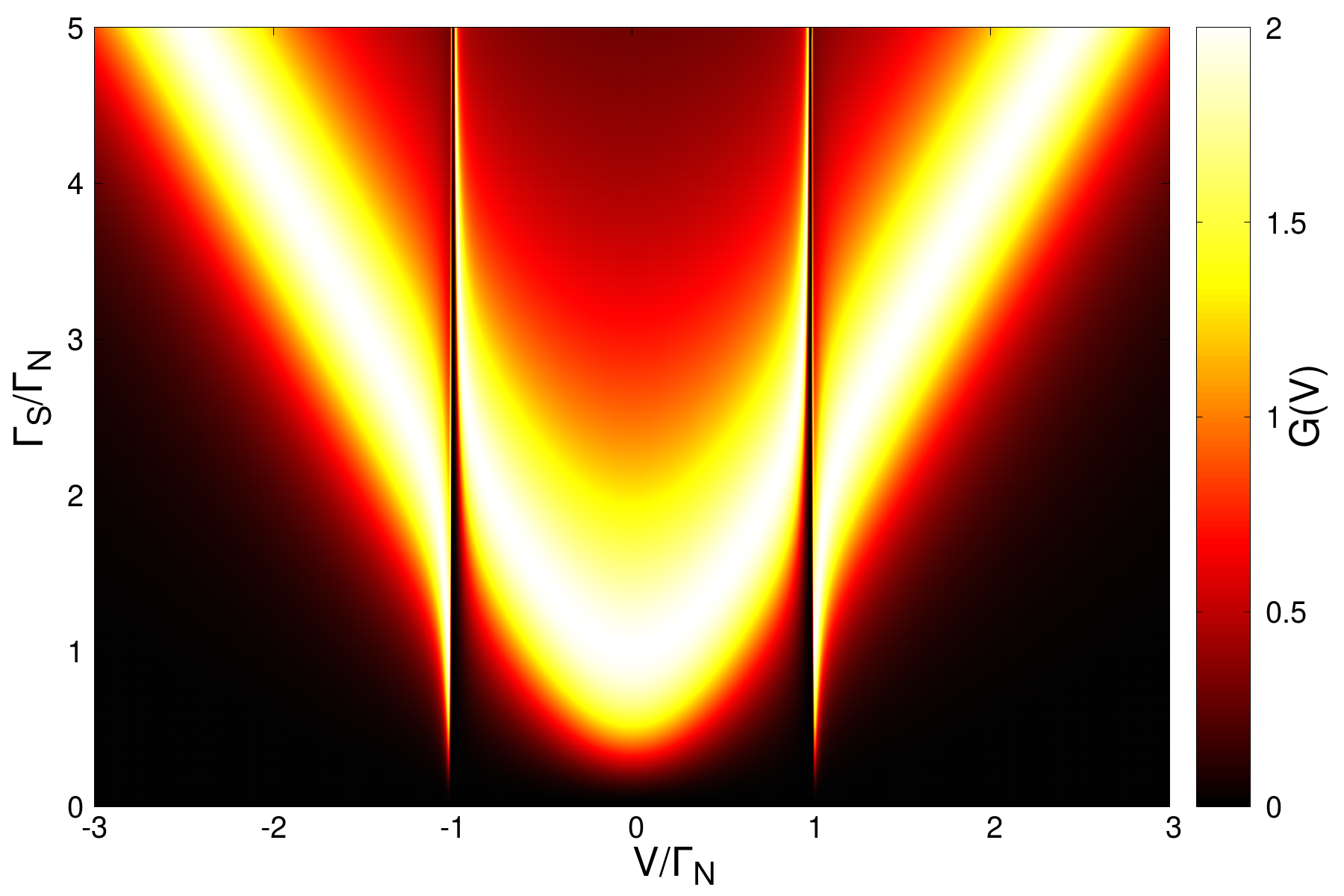}
\caption{The differential Andreev conductance $G(V)$ [in units 
of $2e^{2}/h$] obtained for the same model parameters 
as in Fig. \ref{figure_new}.}
\label{Andreev_cond_free}
\end{figure}

Figure \ref{Andreev_cond_free} shows the differential Andreev 
conductance $G(V)=dI(V)/dV$ obtained for the uncorrelated quantum 
dots. We can notice, that the subgap transport properties are sensitive 
to both the quantum interference (for small $t$) or the Dicke-like effect 
(for the strong interdot coupling). The optimal conductance $4e^{2}/h$ 
occurs at such voltages $V$, which coincide with the subgap quasiparticle 
energies. The Andreev spectroscopy would thus be able to verify the 
aforementioned relationship of the interferometric and/or Dicke 
effect with the proximity induced electron pairing.

\section{Interplay with Kondo effect\label{sec:IV}}

Repulsive interactions $U_{j}$ between opposite spin electrons can 
induce further important effects. It is convenient to describe their 
influence, expressing the matrix Green's function ${\mb G}_{j}(\omega)$ 
via \cite{Bauer-2007,Li-2016}
\begin{equation}
{\mb G}_{j}(\omega)={\mb G}_{j}^{0}(\omega)+
{\mb G}_{j}^{0}(\omega) \; U_{j} \; {\mb F}_{j}(\omega) ,
\label{dressed}
\end{equation}
where ${\mb G}_{j}^{0}(\omega)$ refers to the case $U_{j}=0$, and 
the two-body Green's function ${\mb F}_{j}(\omega)$ is defined as
\begin{equation}
{\mb F}_{j}(\omega)=
\begin{pmatrix}
\langle\langle \hat{d}_{j\uparrow} \hat{n}_{j\downarrow}; 
\hat{d}_{j \uparrow}^{\dagger} \rangle\rangle & 
\langle\langle \hat{d}_{j\uparrow} \hat{n}_{j\downarrow}; 
\hat{d}_{j\downarrow}\rangle\rangle \\
 \langle\langle -\hat{d}_{j\downarrow}^{\dagger} \hat{n}_{j\uparrow} ; 
\hat{d}_{j\uparrow}^{\dagger}\rangle\rangle & 
\langle\langle -\hat{d}_{j\downarrow}^{\dagger}\hat{n}_{j\uparrow} ; 
\hat{d}_{j\downarrow}\rangle\rangle 
\end{pmatrix} .
\end{equation}
In this paper we focus on the correlation effects driven by the potential 
$U_{0}$, because it has the predominant influence on measurable 
transport properties of our system. As concerns $U_{\pm 1}$, they 
could merely mimic the multilevel structure of the side-coupled dots. In 
experimental realizations of the correlated quantum dots coupled to the 
superconducting electrodes \cite{deacon2010,lee2012,pillet2013,Zitko-2015} 
the Coulomb potential $U_{j}$ usually exceeds the superconducting energy 
gap $\Delta$ (at least by one order of magnitude). Under such 
circumstances the correlation effects manifest themselves in the subgap regime
$|\omega|<\Delta$ in a rather peculiar way, via (i) the singlet-doublet 
transition (or crossover) and (ii) the subgap Kondo effect 
\cite{Zitko-2015,Domanski-2016}. 

\subsection{Perturbative approach}

The singlet-doublet transition can be captured already within the lowest 
order (Hartree-Fock-Bogoliubov) decoupling scheme
\begin{equation}
U_{0}{\mb F}_{0}(\omega) \approx
\underbrace{U_{0}
\begin{pmatrix}
\langle \hat{n}_{0\downarrow} \rangle & 
\langle \hat{d}_{0\downarrow} \hat{d}_{0\uparrow} \rangle \\
\langle \hat{d}_{0\uparrow}^{\dagger} \hat{d}_{0\downarrow}^{\dagger}\rangle & 
- \langle \hat{n}_{0\uparrow} \rangle 
\end{pmatrix}}_{\Sigma_{0}^{1st}}
 {\mb G}_{0}(\omega) .
\label{HFB}
\end{equation}
As usually the 1-st order correction (with respect to $U_{0}$) to the selfenergy is static, 
therefore it can be incorporated into the renormalized energy level $\tilde{\epsilon}_{0} 
\equiv \epsilon_{0}+U_{0}\langle \hat{n}_{0\sigma} \rangle$ and the effective 
pairing potential $\tilde{\Gamma}_{S}/2 \equiv \Gamma_{S}/2 - U_{0} \langle 
\hat{d}_{0\downarrow} \hat{d}_{0\uparrow} \rangle$. Such Hartree-Fock-Bogoliubov
corrections (\ref{HFB}) imply a crossing of the subgap Andreev states when the 
ground state changes from the spinful to spinless configuration upon increasing 
the ratio of $\Gamma_{S}/U_{0}$. This effect is known to reverse the tunneling 
current in the Josephson junctions (so called, $0-\pi$ transition) and has been 
extensively studied (see Ref. [\onlinecite{Zonda-2015}] for a comprehensive discussion). 

To describe the subgap Kondo effect it is, however, necessary to go 
beyond the mean field approximation (\ref{HFB}), taking into account the higher 
order (dynamic) corrections
\begin{equation}
U_{0}{\mb F}_{0}(\omega) =
\left[ {\mb \Sigma}_{0}^{1st}+{\mb \Sigma}_{0}^{dyn}(\omega) \right]
 {\mb G}_{0}(\omega) .
\label{dynamic}
\end{equation}
Formally, Eq.\ (\ref{dynamic}) can be recast into the Dyson form 
${\mb G}_{0}(\omega)^{-1}={\mb G}^{0}_{0}(\omega)^{-1}-\left[ 
{\mb \Sigma}_{0}^{1st}+{\mb \Sigma}_{0}^{dyn}(\omega)\right]$. 
Obviously the dynamic part ${\mb \Sigma}_{0}^{dyn}(\omega)$ can 
be estimated  only approximately, because the present problem is 
not solvable.  

In the limit $|\omega|\ll\Delta$ the diagonal and off-diagonal parts 
of the Green's function ${\mb G}_{0}(\omega)$ are interdependent 
through the  (exact) relation \cite{Bauer-2007}
\begin{eqnarray} 
\left( \tilde{\omega} - \epsilon_{0} \right) {\mb G}_{0,11}(\omega) 
= 1 - \frac{\Gamma_{S}}{2} {\mb G}_{0,21}(\omega) 
+U_{0} {\mb F}_{0,11}(\omega) .
\label{exact_relation} 
\end{eqnarray} 
Here $\tilde{\omega}=\omega-\sum_{{\bf k}} \frac{|V_{{\bf k} N}|^{2}}
{\omega\! - \! \xi_{{\bf k} N}}$, which in the wide-band limit simplifies 
to $\tilde{\omega}=\omega+i\Gamma_{N}/2$. We determine the two-body 
propagator ${\mb F}_{0,11}(\omega)=\langle\langle \hat{d}_{0\uparrow}
\hat{n}_{0\downarrow};\hat{d}_{0\uparrow}^{\dagger}\rangle\rangle$ 
using the decoupling scheme within the equation of motion procedure
\cite{EOM-method}
\begin{eqnarray} 
{\mb F}_{0,11}(\omega) \simeq \frac{ \langle 
\hat{n}_{0\downarrow} \rangle - \gamma_{1}(\omega) \; 
{\mb G}_{0,11}(\omega)}{ \tilde{\omega} - \epsilon_{0} 
- U_{0} - \gamma_{3}(\omega)} ,
\label{approximation_for_U}
\end{eqnarray} 
where the auxiliary functions $\gamma_{\nu}(\omega)$ are defined as 
\begin{eqnarray}
\gamma_{\nu}(\omega) &=& \sum_{{\bf k}} 
\!  \left[ \frac{|V_{{\bf k} N}|^{2}}{\omega\! - \! 
\xi_{{\bf k} N} } + \frac{|V_{{\bf k} N}|^{2}}{\omega\! -\! U_{0} \! - 
2 \varepsilon_{0}\!+\!\xi_{{\bf k} N} } \right]  
\nonumber \\
& \times & \left\{ \begin{array}{lc} 
f_{FD}(\xi_{{\bf k}N}) & \mbox{\rm for } \nu=1 ,\\
1 & \mbox{\rm for } \nu=3  . \end{array} \right.   
\label{Kondo_sigma} 
\end{eqnarray}
This method implies the diagonal selfenergy 
${\mb \Sigma}_{0,11}(\omega)={\mb \Sigma}_{0,11}^{1st}+{\mb \Sigma}
_{0,11}^{dyn}(\omega)$  in the familiar form \cite{EOM-method}
\begin{eqnarray} 
\frac{1}{\omega-\epsilon_{0}-{\mb \Sigma}_{0,11}(\omega)}
&=&\frac{1-\langle \hat{n}_{0\downarrow} \rangle}{\tilde{\omega}
-\epsilon_{0}+\frac{U_{0}\gamma_{1}(\omega)}{\tilde{\omega}-\epsilon_{0}
-U_{0}-\gamma_{3}(\omega)}}
\label{sigma_EOM} \\ &+&
\frac{\langle \hat{n}_{0\downarrow} \rangle}{\tilde{\omega}
-\epsilon_{0}-U_{0}+\frac{U_{0}[\gamma_{1}(\omega)-\gamma_{3}(\omega)]}
{\tilde{\omega}-\epsilon_{0}-\gamma_{3}(\omega)}} .
\nonumber 
\end{eqnarray} 
The off-diagonal term ${\mb \Sigma}_{0,21}(\omega)$ can be obtained 
from Eqs. (\ref{exact_relation}) and (\ref{approximation_for_U}). Such
procedure provides the qualitative insight into the Kondo effect, 
spectroscopically 
manifested by the narrow Abrikosov-Suhl peak at $\omega=\mu_{N}$. 

 \begin{figure}
\includegraphics[width=0.9\linewidth]{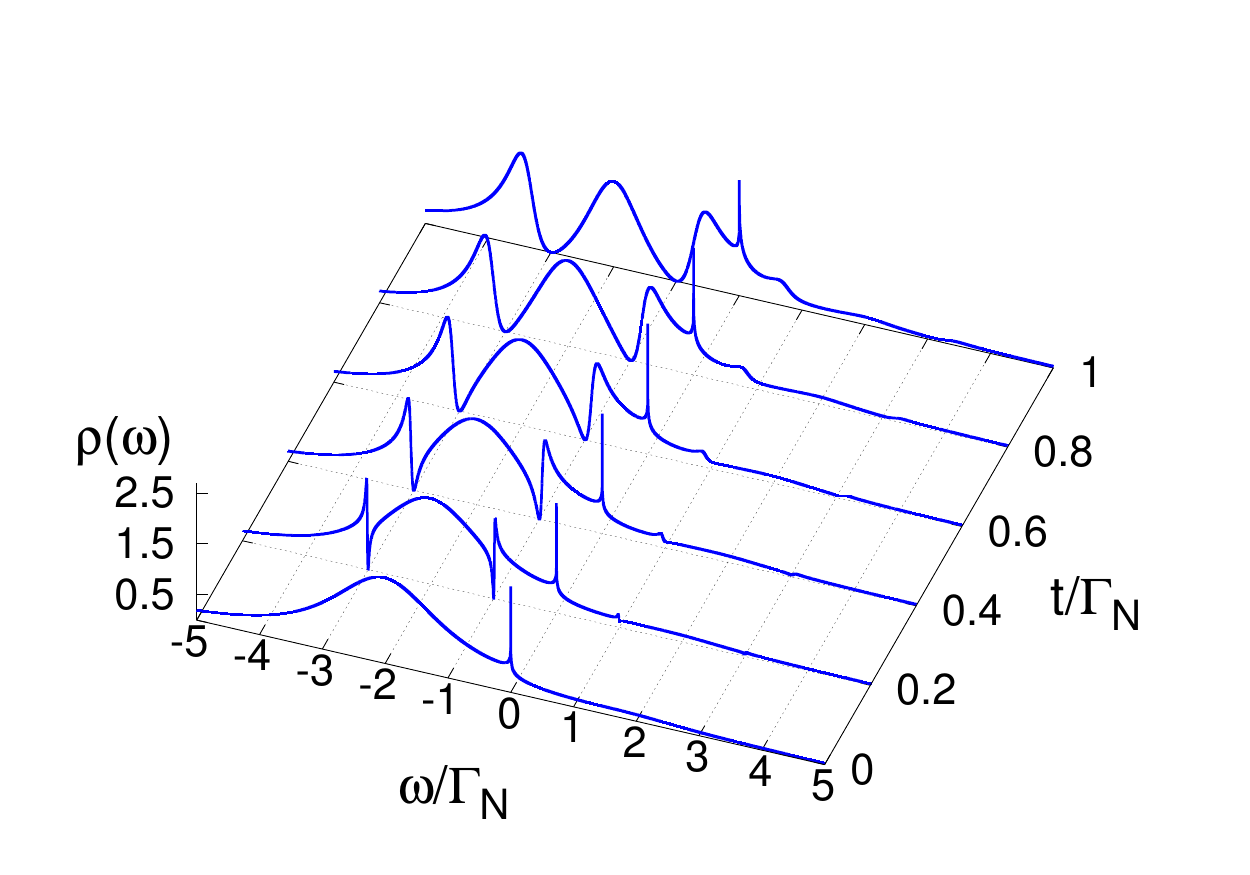}
\caption{Evolution of the spectral function $\rho(\omega)$ obtained for the 
strongly correlated QD$_{0}$ in the Kondo region from the interferometric 
(small $t$) to molecular (large $t$) limits. Calculations have been done 
for $\mu_{N}=0$, $\epsilon_{0}=-2\Gamma_{N}$, $\delta=\Gamma_{N}$, 
$\Gamma_{S}=4\Gamma_{N}$ and $U_{0}=100\Gamma_{N}$.}
\label{from_int_to_Dicke}
\end{figure}

We now investigate the effect of the interdot coupling on the Andreev 
spectroscopy, considering the interferometric and the 
molecular regions. Figure\ \ref{from_int_to_Dicke} shows  
the spectrum of QD$_{0}$ in the Kondo regime at temperature 
$T=10^{-6}\Gamma_{N}$ for $\epsilon_{0}=-2\Gamma_{N}$, $\delta=
\Gamma_{N}$, $\Gamma_{S}=4\Gamma_{N}$, assuming the large Coulomb potential 
$U_{0}=100\Gamma_{N}$. Initially, for $t=0$, the spectral function $\rho(\omega)$ reveals: 
(i) the quasiparticle peak at $\omega \approx \epsilon_{0}$, (ii) its tiny particle-hole  
companion at $\omega \approx - \epsilon_{0}$ (let us remark that superconducting 
proximity effect substantially weakens upon increasing $|\epsilon_{0}|/\Gamma_{S}$) 
and (iii) the narrow Abrikosov-Suhl peak at $\omega=\mu_{N}$ (manifesting the Kondo 
effect). For the weak interdot coupling $t\ll\Gamma_{N}$, we notice appearance of the 
Fano-type (interferometric) features at $\omega=\epsilon_{\pm 1}$. For the stronger 
coupling $t$, the spectrum of QD$_{0}$ evolves to its molecular-like structure, 
resembling the one discussed in the preceding section. Upon increasing $t$,
the subradiant quasiparticle (centered around $\epsilon_{0}$) gradually narrows, 
whereas the superradiant quasiparticles absorb more and more spectral weights. 
Such transfer of the spectral weight indirectly amplifies the Abrikosov-Suhl 
peak, existing on the upper superradiant quasiparticle. 

\begin{figure}
\includegraphics[width=0.9\linewidth]{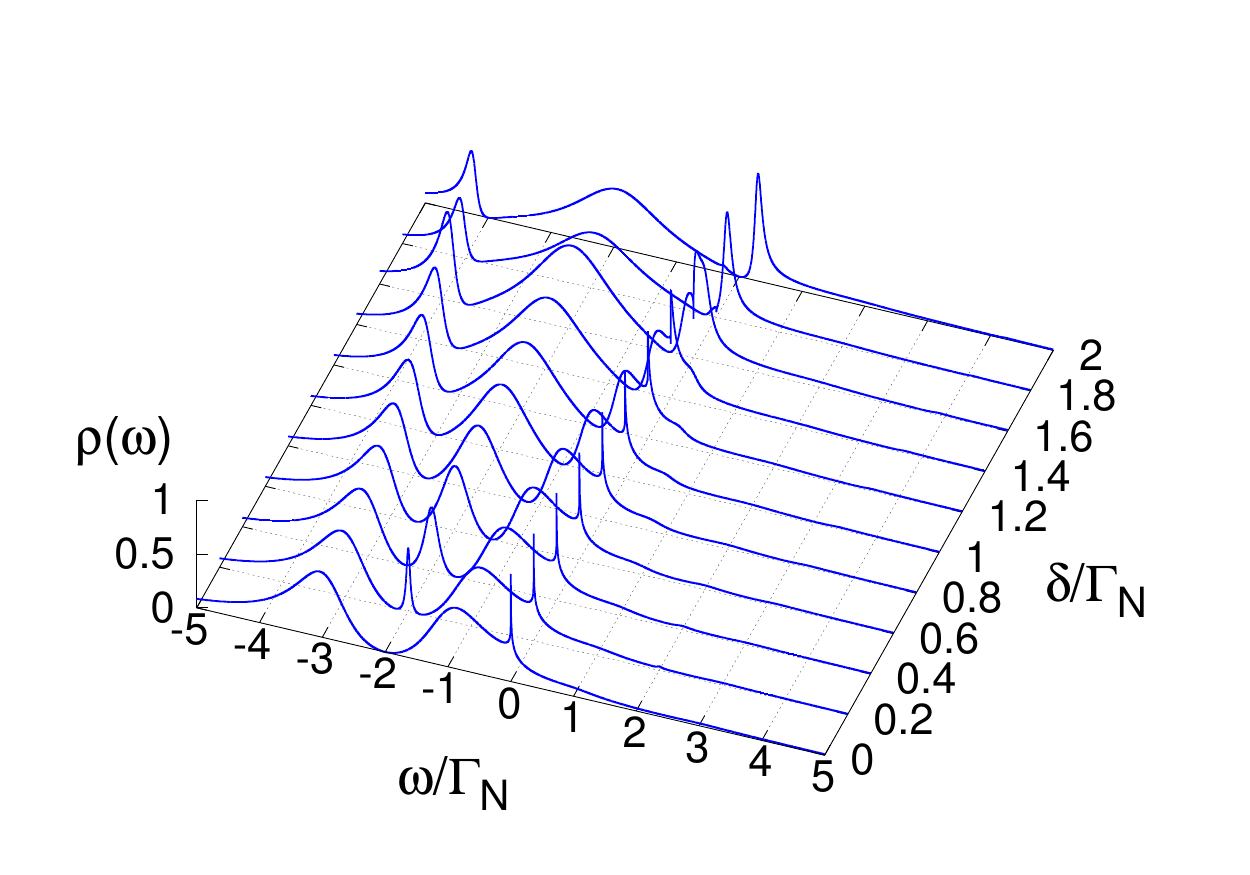}
\caption{Spectral function of the correlated central quantum dot
in the Kondo regime obtained for $t=2\Gamma_{N}$ using 
$\epsilon_{0}=-2\Gamma_{N}$, $\Gamma_{S}=4\Gamma_{N}$, 
$U_{0}=100\Gamma_{N}$.}
\label{detuning_effect}
\end{figure}

In the discussed case the Dicke effect constructively amplifies the Abrikosov-Suhl 
peak, but in general the Kondo effect can depend on the detuning $\delta$. This
is illustrated in Fig.\ \ref{detuning_effect}, where upon varying $\epsilon_{\pm 1}
-\epsilon_{0}$ the Abrikosov-Suhl peak is  enhanced up to some critical 
detuning $\delta_{crit} \sim t$, at which destructive interference depletes 
all the electronic states near $\mu_{N}$.

\begin{figure}
\includegraphics[width=0.95\linewidth]{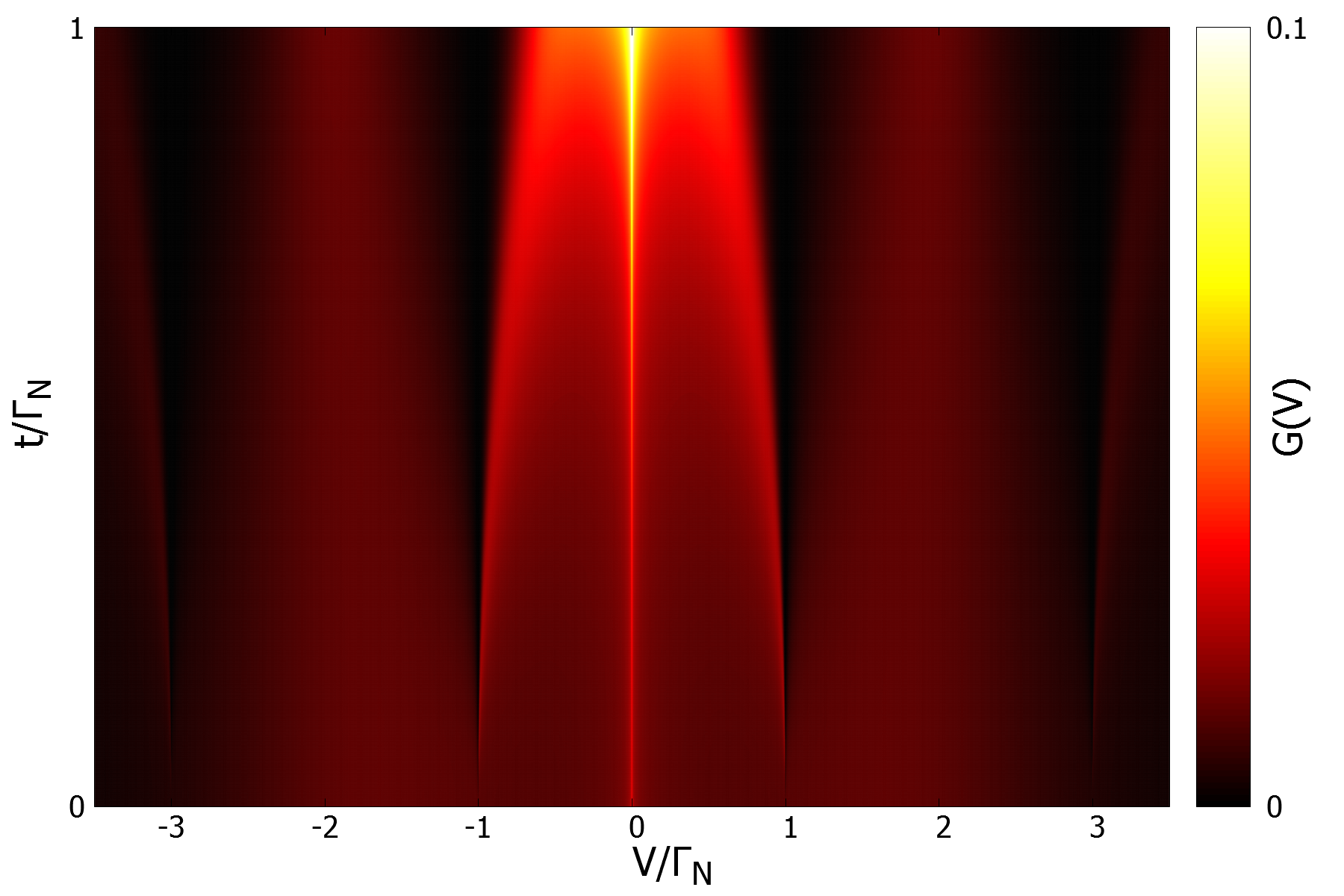}
\caption{The differential Andreev conductance $G(V)=dI(V)/dV$ 
[in units $2e^{2}/h$] as a function of the interdot coupling 
$t$, ranging from the weak (interferometric) to molecular 
(Dicke) regions. Calculations have been done for the same 
model parameters as in Fig. \ref{from_int_to_Dicke}.}
\label{cond_vs_t}
\end{figure}

In Fig.\ \ref{cond_vs_t} we show the differential Andreev conductance 
obtained for our setup at temperature $T=10^{-6}\Gamma_{N}$ as a function 
of the voltage $V$ and the interdot coupling $t$. In the absence of the side-attached 
dots ($t=0$) we notice two broad maxima at $|eV|\approx \epsilon_{0}$ (corresponding 
to energies of the subgap quasiparticle states) and the zero-bias peak (due to the Kondo 
effect). For finite and weak interdot coupling $t\ll\Gamma_{N}$, the quantum 
interference starts to play a role as manifested by the asymmetric Fano-type 
resonances at $\epsilon_{\pm 1}$. With further increase of $t$ we observe  
development of the sub- and superradiant features, typical for the molecular 
regime. Transfer of the spectral weight from the subradiant to superradiant 
states amplifies the zero-bias conductance (bright region at $V\sim 0$). 

\begin{figure}
\includegraphics[width=0.95\linewidth]{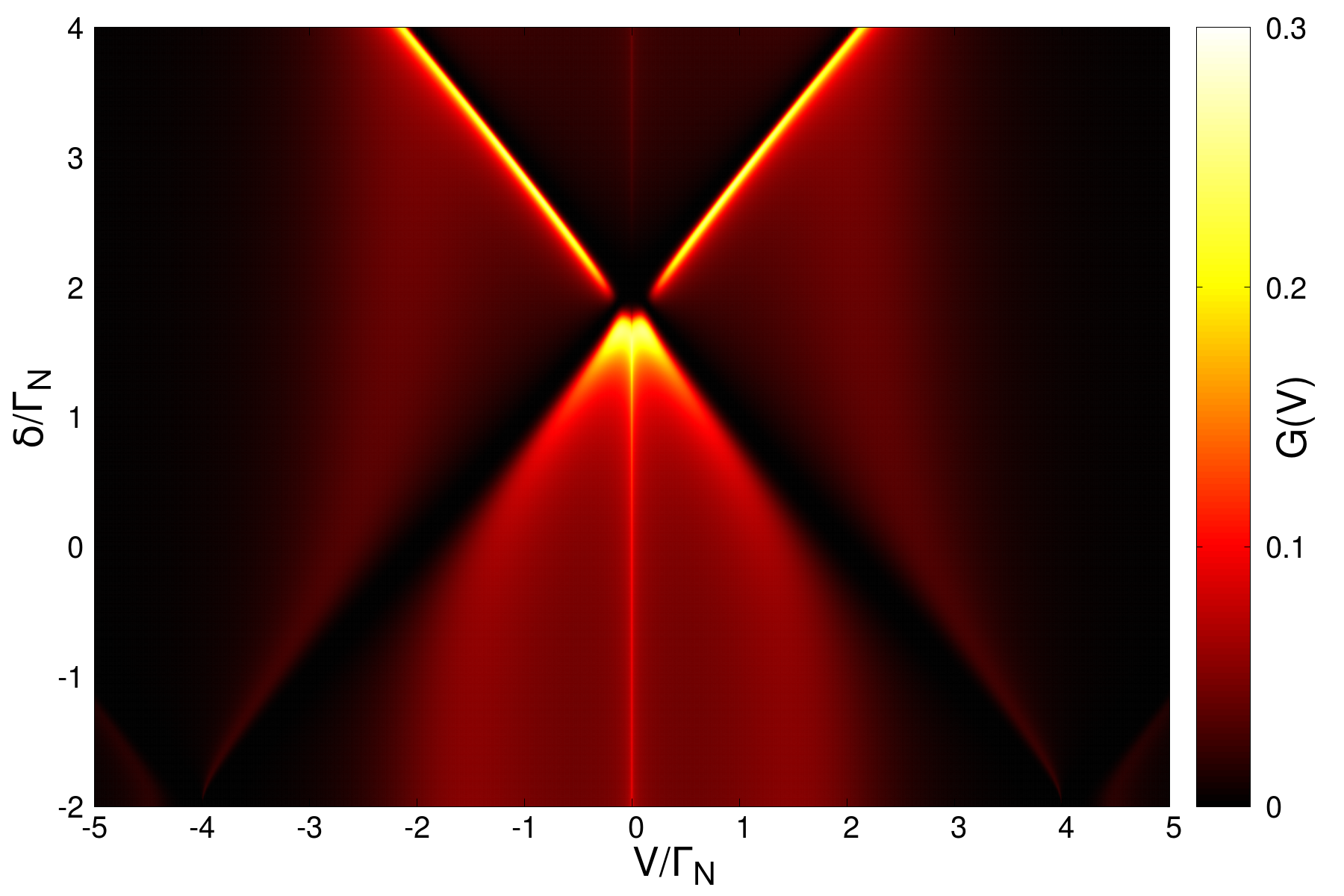}
\caption{The differential Andreev conductance $G(V)$ versus
$\delta$ obtained for the same set of parameters
used in Fig. \ref{detuning_effect}.}
\label{cond_vs_delta}
\end{figure}

Figure \ref{cond_vs_delta} shows the evolution of the differential Andreev conductance 
with respect to $\delta$ for the same set of parameters as used in Fig.\ 
\ref{detuning_effect}. Since the zero-bias conductance probes the quasiparticle 
states at $\omega\sim 0$, it tells us (indirectly) about behavior of the 
subgap Kondo effect. The ongoing redistribution of the spectral weight between 
the subradiant and superradiant states enhances this zero-bias conductance 
until the critical detuning $\delta_{c} \approx t$. Above this critical 
detuning, the Kondo effect is completely washed out, signaling qualitative 
change of the QD$_{0}$ ground state. For a better understanding of the low 
energy physics, we perform nonperturbative calculations based on the numerical 
renormalization group (NRG) technique. 

\subsection{NRG results}

For a reliable analysis of a subtle interplay between the correlations, 
the induced electron pairing, and the sub/super-radiant Dicke states we 
performed the numerical renormalization group calculations \cite{Wilson}. 
Our major concern was to investigate the low energy Kondo physics 
appearing in the sub-gap regime due to the spin-exchange interactions 
between the central quantum dot and the metallic lead \cite{Domanski-2016}. 
In such deep subgap regime (\ref{proximized_QD0}) the quantum dot hybridized 
with the superconducting reservoir can be described by the effective 
Hamiltonian \cite{Bauer-2007,RozhkovArovas}
\begin{eqnarray}
\hat{H}_{QD} + \hat{H}_{S} + \hat{H}_{QD-S} \rightarrow
\hat{H}_{QD} - \frac{\Gamma_S}{2}\left( \hat{d}^\dagger_{0\uparrow}
\hat{d}^\dagger_{0\downarrow} + \hat{d}_{0\downarrow} 
\hat{d}_{0\uparrow} \right).
\nonumber
\end{eqnarray}
Under such conditions the initial Hamiltonian (\ref{model}) simplifies 
to the single-channel model, allowing for a vast reduction of computation 
efforts. We performed NRG calculations, using the Budapest Flexible DM-NRG 
code \cite{fnrg} for constructing the density matrix 
of the system\cite{AndersSchiller,dm-nrg} and determining the matrix Green's 
function (\ref{GF_definition}). During the calculations we exploited the 
spin SU(2) symmetry and kept $N_{\rm kept}=3000$ multiplets. We obtained 
the satisfactory solution within $N=50$ iterative steps, assuming a flat 
density of states of the normal lead with a cutoff $D=U_{0}$ and imposing
the discretization parameter $\Lambda=2$. To improve the quality
of the spectral data, our results were averaged over 
$N_z=4$ interleaved discretizations \cite{Z}. Next, we
determined the real parts of $\mb{G}(\omega)$
[needed for the Andreev transmittance] from the Kramers-Kr\"onig relations.

\begin{figure}
\includegraphics[width=0.95\linewidth]{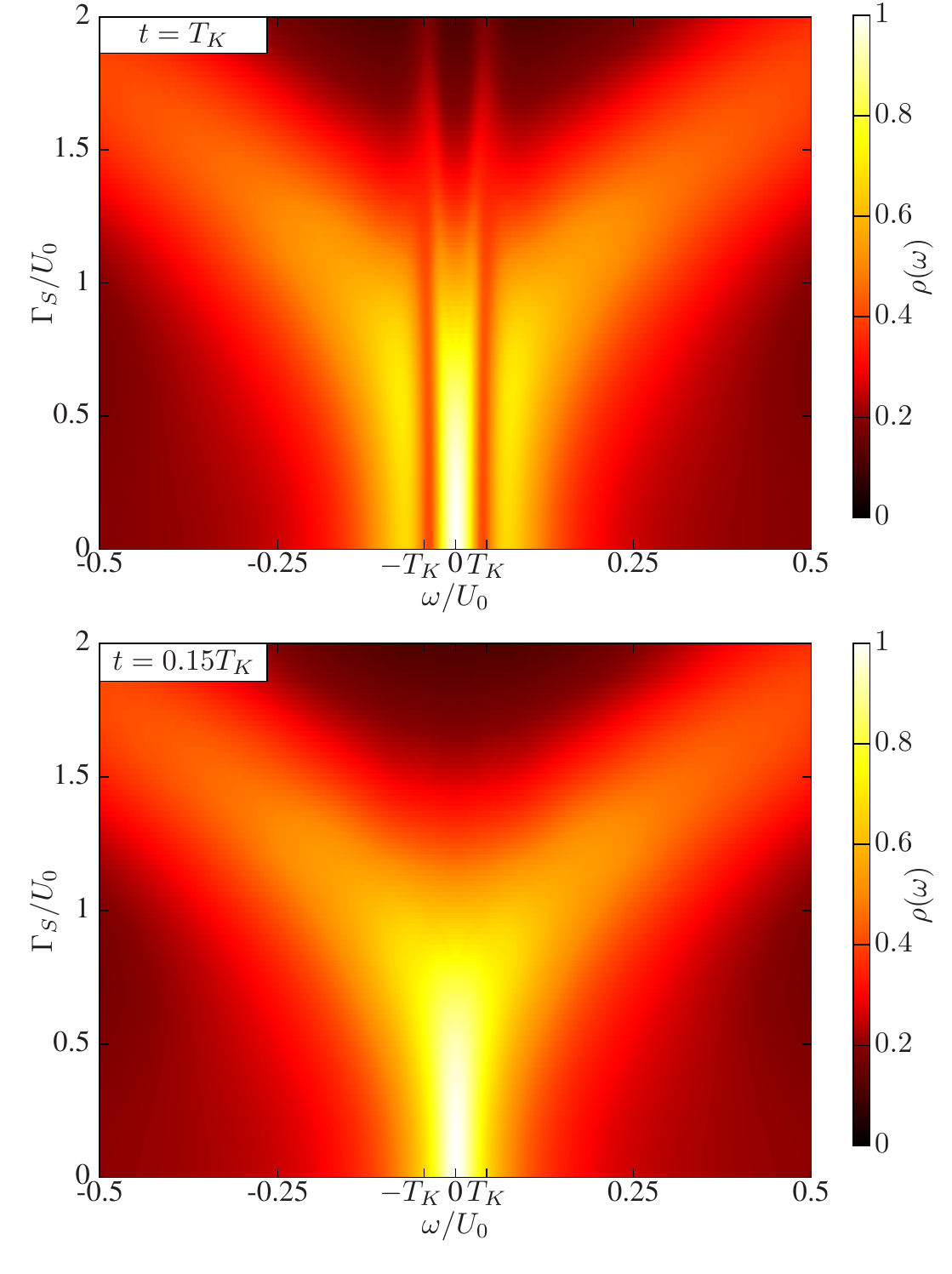}
\caption{The spectral function of the half-filled QD$_{0}$ obtained by NRG 
        for $\Gamma_{N}=0.4U_0$, $\delta = T_K = 0.044U_0$, $T=0$, and 
        $t=\delta$ (top panel) or $t=0.15\delta$ (bottom panel).}
\label{NRG-A}
\end{figure}

Figure~\ref{NRG-A} presents the spectral function obtained by NRG for QD$_{0}$ using 
$\Gamma_{N}=0.4U_0$ and $\delta = T_K = 0.044U_0$, where the Kondo temperature $T_K$ 
is estimated with the Haldane formula \cite{Haldane} in the case of $t=\Gamma_S=0$.
For $t=\delta$, some similarities to the bottom panel of Fig.~\ref{figure_new} 
can be observed. First of all, for $\Gamma_S=0$, the spectral function $\rho(\omega)$
exhibits a peak at $\omega=0$ and two side peaks at frequencies $\omega \approx \pm \delta$.
When $\delta=T_K$, the side peaks are very close to the central one and are definitely
less sharp than for the non-interacting case presented in Fig.~\ref{figure_new}. 
For $\Gamma_S \gtrsim U_0$, the Abrikosov-Suhl peak smoothly evolves into the 
Andreev quasiparticle states \cite{Domanski-2016} and the spectral weight 
is successively shifted towards the side peaks. For stronger coupling $\Gamma_{S}$,
the Kondo effect is no longer present.

For $t=0.15T_K$, the situation is rather different (bottom panel of 
Fig.~\ref{NRG-A}) because instead of the Dicke effect we can see only some
interferometric signatures. For small $\Gamma_S$, the spectral function 
is characterized by the single Abrikosov-Suhl peak. With increasing $\Gamma_{S}$ 
such peak gradually broadens, and finally for $\Gamma_S \gtrsim U_0$ it splits 
because of a quantum phase transition from the spinful (doublet) to the spinless 
(singlet) configurations \cite{Bauer-2007,Domanski-2016}. We presume that 
the inter-dot coupling $t=0.15T_K$ is too weak to have any significant influence 
on the low-energy properties of our system  (unlike the case considered in the 
bottom panel of Fig.~\ref{figure_new}). Yet, the spectral weight transfer 
towards the higher energies with increasing $\Gamma_{S}$ is quite evident.

\begin{figure}
\includegraphics[width=0.95\linewidth]{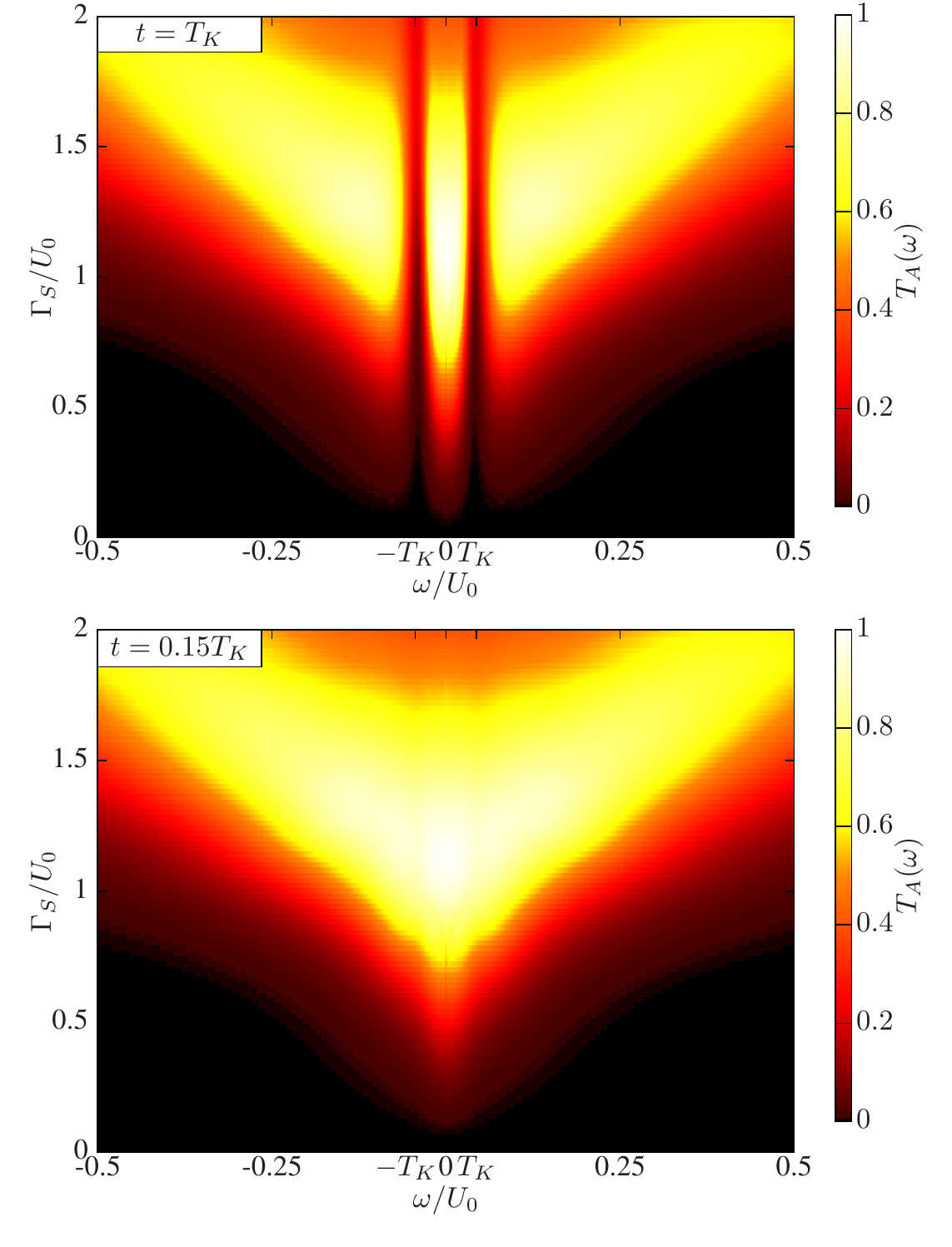}
\caption{The Andreev transmittance  $T_{A}(\omega)$ obtained by NRG  
         for the set of model parameters corresponding to Fig.~\ref{NRG-A}.}
\label{NRG-TA}
\end{figure}

The observations shown in Fig.~\ref{NRG-A} have their consequences for the 
measurable transport quantities. Results for the zero-temperature Andreev transmittance 
$T_A(\omega)$ obtained by the NRG calculations are presented in 
Fig.~\ref{NRG-TA}. At zero temperature, the Andreev transmittance 
has a simple relationship with the differential conductance
$G(V)= \frac{2e^{2}}{h} [T_{A}(\omega=eV) + T_{A}(\omega=-eV)]$. 
For small $\Gamma_{S}$, the energetically favorable ground state configuration 
of QD$_{0}$ is $\left| \sigma \right>$, therefore it is hardly affected by 
the superconducting proximity effect, hence the Andreev transmittance [dependent
on the off-diagonal terms of the matrix Green's function $\mb{G}_{0}(\omega)$] 
is negligibly small. With increasing $\Gamma_S$ the central quantum dot
evolves to the BCS-type configuration $v\left| 0 \right> - u \left| 
\uparrow\downarrow\right>$, therefore efficiency of the pairing effects
is significantly enhanced as can be seen by bright areas in Fig.~\ref{NRG-TA} 
for $\Gamma_S \gtrsim U_0$.  Such changeover of the QD$_{0}$ ground state is,
however, detrimental to the Kondo effect because the spinless BCS-type 
configuration cannot be screened. For  $\Gamma_{N}\neq 0$, this 
quantum phase transition is a crossover, therefore the Abrikosov-Suhl 
peak (present at $\omega=0$ for $\Gamma_{S} < U_{0}$) evolves in a fuzzy 
manner into the Andreev quasiparticles (existing at finite energies). More detailed
description of this mechanism has been previously discussed (for the single
quantum dot heterostructure) by several authors \cite{Bauer-2007,Domanski-2016}.

Let us remark that for $t=T_K$ (top panel in Fig.~\ref{NRG-TA}) the subgap 
transport properties can clearly distinguish between the subradiant and 
superradiant contributions. Since the Kondo effect is very much affected 
by the induced electron pairing, its interplay with the Dicke effect becomes 
highly nontrivial. Empirical observability of the subgap Kondo effect would 
be, however, feasible only when approaching the singlet to doublet 
crossover (i.e.\ when $\Gamma_S \sim U_{0}$). This fact is unique and it 
has no resemblance to the properties of triple quantum dots embedded between 
the normal metallic leads.

\section{Summary}

We have studied  nontrivial interplay between the proximity induced electron 
pairing and the Dicke-like effect in a heterojunction, comprising three quantum 
dots vertically coupled between the normal and superconducting leads. This
setup allows for a smooth evolution from the weak inter-dot coupling regime,
characterized by the Fano-type interferometric features, to the strong 
coupling (or `molecular') region, revealing signatures of the Dicke-like 
effect even in absence of correlations. In the latter case the narrow 
(subradiant) and the broad (superradiant) contributions can be formed 
either by (i) increasing the interdot coupling $t$ or (ii) reducing the 
detuning  $\delta$ of their energies \cite{trocha2008a,trocha2008b,vernek2010}. 
We have examined the electronic structure of central quantum dot, finding 
transfer of its spectral weights from the low- to the high-energy 
states caused by the induced electron pairing.

In the weak inter-dot coupling (interferometric) regime, the usual 
subagp quasiparticles (Andreev states) are superimposed with the 
Fano-type resonant lineshapes appearing at $\omega=\epsilon_{\pm 1}$. 
In the molecular region (for large $t$), 
the sub- and superradiant states undergo the splitting. Since the 
subradiant state is restricted to the energy region $\omega \in 
\left( \epsilon_{-1},\epsilon_{+1} \right)$, its splitting is 
bounded from above. For this reason the strong  electron pairing 
is detrimental for it, transferring the spectral weight towards the 
superradiant states. Influence of the electron pairing on the subradiant 
state can indirectly amplify the subgap Kondo effect (provided that 
$\mu_{N} < \epsilon_{-1}$ or $\mu_{N} > \epsilon_{+1}$) shown 
by enhancement of the zero-bias Andreev conductance.

We also examined the rich interplay between the correlations, electron pairing 
and  influence of the side-attached quantum dots by the perturbative method 
and using the NRG
technique. In particular, we argue that the Kondo-Dicke features
would be empirically observable only near the singlet-doublet quantum 
phase transition (crossover). Such subtle effect is caused by crossing
of the subgap Andreev quasiparticles which is accompanied by qualitative 
changeover between the different ground state configurations. The Dicke
effect is restricted exclusively to the spinful (doublet) regime, 
which  for the half-filled central quantum dot occurs when 
$ \Gamma_{S} \gtrsim U_{0}$. Such complicated many-body effects can be 
experimentally probed by the subgap Andreev spectroscopy.

\section*{Acknowledgments}

This work is supported by the National Science Centre (Poland) 
via the projects DEC-2014/13/B/ST3/04451 (SG, TD) and
DEC-2013/10/E/ST3/00213 (KPW, IW).
Computing time at Pozna\'n Superconducting
and Networking Center is acknowledged.

\appendix
\section{Normal heterostructures}

\begin{figure}
\includegraphics[width=0.9\linewidth]{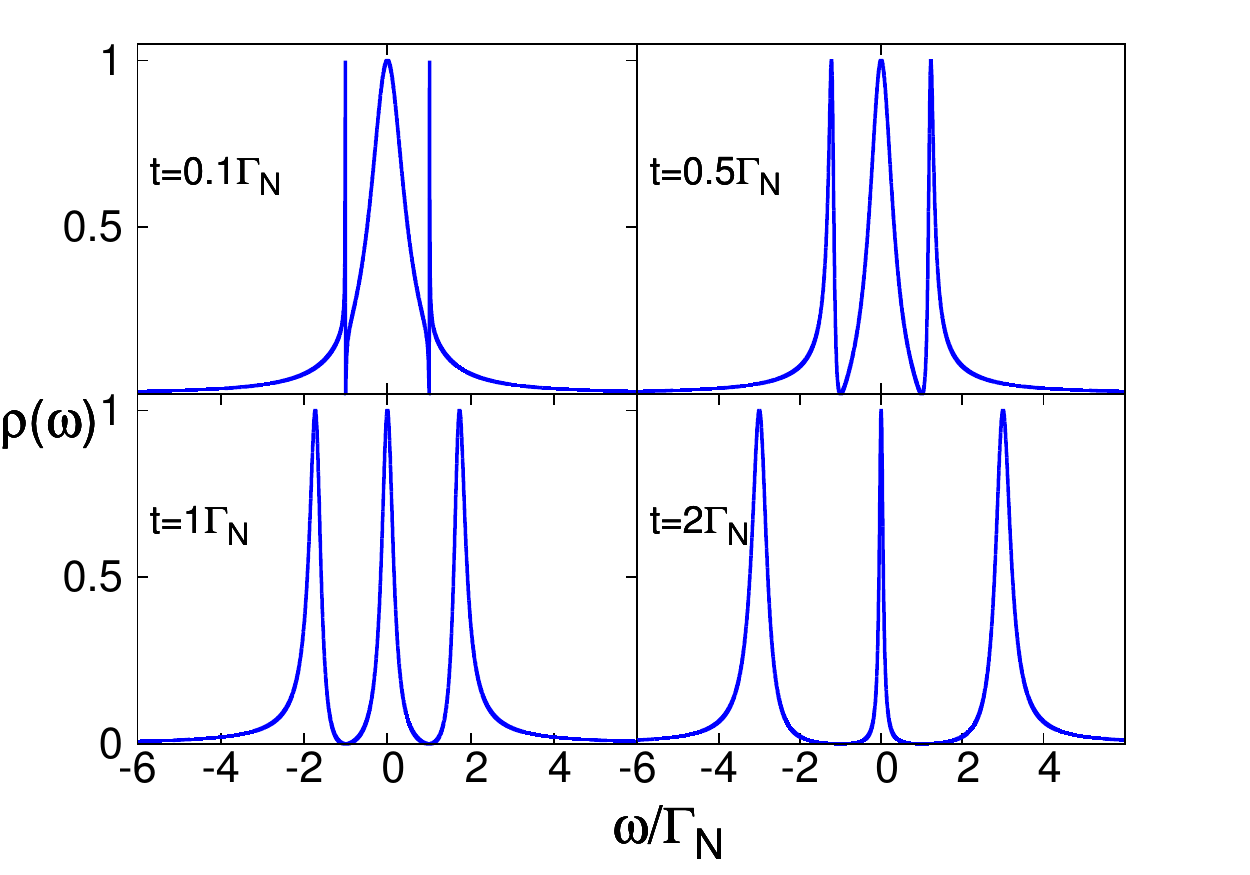}
\caption{Evolution of the spectral function $\rho(\omega)$ for
varying interdot coupling, from the interferometric
(small $t$) to the molecular (large $t$) regions.}
\label{from_Fano_to_Dicke_normal}
\end{figure}

In this appendix we illustrate how the molecular (i.e.\ three peak 
structure of QD$_{0}$ spectrum) gradually emerges from the interferometric 
(weak interdot coupling) scenario, considering both the external electrodes 
to be metallic \cite{trocha2008a,trocha2008b,vernek2010}. For simplicity we 
neglect the correlations and introduce the effective coupling $\Gamma_{N}
+\Gamma_{S} \rightarrow \Gamma_{N}$. The selfenergy is diagonal, therefore 
we can restrict our considerations only to the `11' term
\begin{equation}
{\mb \Sigma}^{U_{0}=0}_{0,11}(\omega)= -i\frac{\Gamma_{N}}{2} +
\frac{t^2}{\omega-\epsilon_{+1}} + 
\frac{t^2}{\omega-\epsilon_{-1}}.
\end{equation}

\begin{figure}
\includegraphics[width=0.9\linewidth]{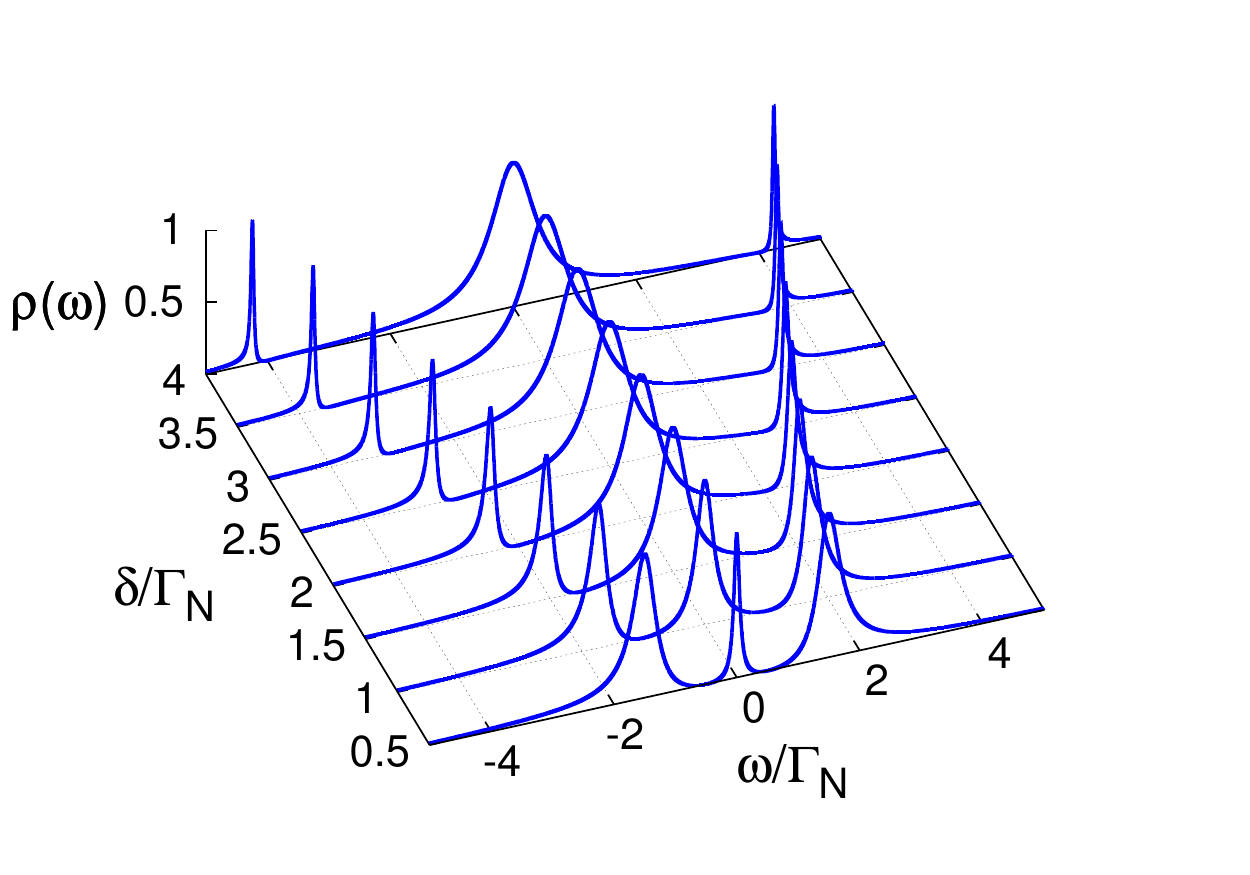}
\caption{Evolution of the spectral function  $\rho(\omega)$ with 
respect to the detuning energy $\delta$ obtained for the uncorrelated 
case ($U_{j}=0$) in the normal heterostructure ($\Delta=0$).}
\label{normal_Dicke_effect}
\end{figure}

Figure \ref{from_Fano_to_Dicke_normal} displays the spectral function 
$\rho(\omega)$ calculated for several values of $t$. For small values 
of the interdot coupling, the QD$_{0}$ spectrum reveals the asymmetric 
Fano-type lineshapes \cite{fano1961} at $\omega \epsilon_{\pm 1}$. Such 
structures arise when a dominant (broad) transport channel interferes 
with a discrete (narrow) state, and can be realized in many areas of 
physics \cite{miroshnichenko2010}. In our case, the Fano resonances originate 
by combining a ballistic transport through the central QD$_{0}$ with additional 
pathways to/from the adjacent QD$_{\pm 1}$. 
By increasing $t$, the Fano resonances gradually smoothen, and all electronic 
states nearby the QD$_{\pm 1}$ levels $\epsilon_{\pm 1}$ are effectively depleted.
In consequence, this induces the three peak (molecular) structure reported 
in the previous studies \cite{vernek2010}. Further increase of the interdot
coupling causes a transfer of the spectral weight from the central (subradiant) 
to the satellite (superradiant) quasiparticle states.

Appearance of the narrow (subradiant) and the broad (superradiant) quasiparticle 
states can be also induced for a fixed interdot coupling $t$, by reducing 
the detuning energy $\delta$. This behavior is shown in Fig.\ 
\ref{normal_Dicke_effect}.


\begin{thebibliography}{11}

\bibitem{Orellana2006}
P.A.\ Orellana, G.A.\ Lara, and E.V.\ Anda,
{\em Kondo and Dicke effect in quantum dots side coupled to a quantum wire},
Phys. Rev. B {\bf 74}, 193315 (2006).

\bibitem{Dicke1953}
R.H.\ Dicke,
{\em The effect of collisions upon the Doppler width of spectral lines},
Phys. Rev. {\bf 89}, 472 (1953).

\bibitem{Raikh1994}
T.V.\ Shahbazyan and M.E.\ Raikh,
{\em Two-channel resonant tunneling},
Phys. Rev. B {\bf 49}, 17123 (1994). 

\bibitem{Wunsch2003}
B.\ Wunsch and A.\ Chudnovskiy,
{\em Quasistates and their relation to the Dicke effect in a mesoscopic ring coupled to a reservoir},
Phys. Rev. B {\bf 68}, 245317 (2003).

\bibitem{Vorrath2003}
T.\ Vorrath and T.\ Brandes,
{\em Dicke effect in the tunnel current through two double quantum dots},
Phys. Rev. B {\bf 68}, 035309 (2003).

\bibitem{Orellana2004}
P.A.\ Orellana, M.L.\ Ladr\'on de Guevara, and F.\ Claro,
{\em Controlling Fano and Dicke effects via a magnetic flux in a two-site Anderson model},
Phys. Rev. B {\bf 70}, 233315 (2004).

\bibitem{Brandes2005}
T.\ Brandes, 
{\em Coherent and collective quantum optical effects in mesoscopic systems},
Phys. Rep. {\bf 408}, 315 (2005).


\bibitem{guevara2006}
M.L.\ de Guevara and P.A.\ Orellana, 
{\em Electronic transport through a parallel-coupled triple quantum 
dot molecule: Fano resonances and bound states in the continuum}, 
Phys. Rev. B {\bf 73}, 205303 (2006).

\bibitem{trocha2008a}
P.\ Trocha and J. Barna\'s,
{\em Dicke-like effect in spin-polarized transport through coupled quantum dots}, 
J. Phys.: Condens. Matt. {\bf 20}, 125220 (2008).

\bibitem{trocha2008b}
P.\ Trocha and J.\ Barna\'s, 
{\em Kondo-Dicke resonances in electronic transport through triple quantum dots}, 
Phys. Rev. B {\bf 78}, 075424 (2008).

\bibitem{vernek2010}
E.\ Vernek, P.A.\ Orellana, and S.E.\ Ulloa, 
{\em Suppression of Kondo screening by the Dicke effect in multiple quantum dots}, 
Phys. Rev. B {\bf 82}, 165304 (2010).

\bibitem{baruselli2013}
P.P.\ Baruselli, R.\ Requist, M.\ Fabrizio, and E.\ Tosatti, 
{\em Ferromagnetic Kondo effect in a triple quantum dot system}, 
Phys. Rev. Lett. {\bf 111}, 047201 (2013).

\bibitem{wang2013}
Q.\ Wang, H.\ Xie, Y.-H.\ Nie, and W.\ Ren, 
{\em Enhancement of thermoelectric efficiency in triple quantum dots 
by the Dicke effect}, 
Phys. Rev. B {\bf 87}, 075102 (2013).








\bibitem{bai2010a}
L.\ Bai, R.\ Zhang, and C.-L.\ Duan, 
{\em Andreev reflection tunneling through a triangular triple quantum dot system}, 
Physica B {\bf 405}, 4875 (2010).

\bibitem{bai2010}
L.\ Bai, Y.-J.\ Wu, and B.\ Wang, 
{\em Andreev reflection in a triple quantum dot system coupled 
with a normal-metal and a superconductor}, 
phys. stat. sol. (b) {\bf 247}, 335 (2010).

\bibitem{ye2015}
C.-Z.\ Ye, W.-T. Lu, H.\ Jiang, and C.-T. Xu,
{\em The Andreev tunneling behaviors in the FM/QD/SC system with intradot spin-flip scattering},
Physica E {\bf 74}, 588 (2015).

\bibitem{xu2016}
W.-P.\ Xu, Y.-Y.\ Zhang, Q.\ Wang, Z.-J.\ Li, and Y.-H.\ Nie, 
{\em Thermoelectric effects in triple quantum dots coupled to 
a normal and a superconducting leads}, 
Phys. Lett. A {\bf 380}, 958 (2016).


\bibitem{yi2016}
G.-Y.\ Yi, X.-Q.\ Wang, H.-N.\ Wu, and W.-J.\ Gong, 
{\em Nonlocal magnetic configuration controlling realized in 
a triple-quantum-dot Josephson junction}, 
Physica E {\bf 81},  26 (2016).

\bibitem{wang2016}
X.-Q.\ Wang, G.-Y.\ Yi,  and W.-J.\ Gong, 
{\em Dicke-Josephson effect in a cross-typed triple-quantum-dot junction}, 
Sol. State Commun. {\bf 247}, 12 (2016).



\bibitem{Bauer-2007}
J.\ Bauer, A.\  Oguri,  \& A.C.\ Hewson,  
{\em Spectral properties of locally correlated electrons in a Bardeen-Cooper-Schrieffer superconductor},
J.\ Phys.: Condens.\ Matter {\bf 19}, 486211 (2007).

\bibitem{Li-2016}
L.\ Li, Z.\ Cao, T.-F.\ Fang, H.-G.\ Luo, and W.-Q.\ Chen,
{\em Kondo screening of Andreev bound states in a normal-quantum dot-superconductor system},
Phys.\ Rev.\ B {\bf 94}, 165144 (2016).

\bibitem{deacon2010}
R.S.\ Deacon, Y.\ Tanaka, A.\ Oiwa, R.\ Sakano, K.\ Yoshida, K.\ Shibata, 
K.\ Hirakawa, and S.\ Tarucha, 
{\em Tunneling spectroscopy of andreev energy levels in a quantum dot coupled 
to a superconductor}, 
Phys. Rev. Lett. {\bf 104}, 076805 (2010).

\bibitem{lee2012}
E.J.H.\ Lee, X.\ Jiang, R.\ Aguado, G.\ Katsaros, C.M.\ Lieber, and S.\ De Franceschi, 
{\em Zero-bias anomaly in a nanowire quantum dot coupled to superconductors},
Phys. Rev. Lett. {\bf 109}, 186802 (2012).

\bibitem{pillet2013}
J.-D.\ Pillet, P.\ Joyez, R\ \v{Z}itko, and M.F. Goffman, 
{\em Tunneling spectroscopy of a single quantum dot coupled to 
a superconductor: from Kondo ridge to Andreev bound states}, 
Phys. Rev. B {\bf 88}, 045101 (2013).

\bibitem{Zitko-2015}
   R.\ \v{Z}itko, J.S.\ Lim, R.\ L\'opez, and R.\ Aguado,
{\em Shiba states and zero-bias anomalies in the hybrid normal-superconductor Anderson model}, 
  Phys.\ Rev.\ B {\bf 91}, 045441 (2015).

\bibitem{Domanski-2016}
T.\ Doma\'nski, I.\ Weymann, M.\ Bara\'nska, and G.\ G\'orski,
{\em Constructive influence of the induced electron pairing on the Kondo state},
Sci.\ Rep.\ {\bf 6}, 23336 (2016).

\bibitem{Zonda-2015}
M.\ \v{Z}onda,   V.\  Pokorn\'y, V.\ Jani\v{s},  \&  T.\ Novotn\'y, 
{\em Perturbation theory of a superconducting $0$ - $\pi$ impurity quantum phase transition},
Sci.\ Rep.\ {\bf 5}, 8821 (2015);
T.\ Doma\'nski, M.\ \v{Z}onda, V. Pokorn\'y, G.\ G\'orski, V. Jani\v{s}, T. Novotn\'y,
{\em Josephson-phase-controlled interplay between the correlation 
effects and electron pairing in a three-terminal nanostructure}
Phys.\ Rev.\ B {\bf 95}, 045104 (2017).

\bibitem{EOM-method}
H.J.W.\ Haug and A.-P.\ Jauho, 
{\em Quantum kinetics in transport and optics of semiconductors},  
Springer-Verlag, (Berlin, Heidelberg, New York) (2008).

\bibitem{Wilson}
	K. G. Wilson, 
	{\em The renormalization group: Critical phenomena and the Kondo problem},
	Rev. Mod. Phys. \textbf{47}, 773 (1975).

\bibitem{RozhkovArovas}
	A.V. Rozhkov and D.P. Arovas, 
	{\em Interacting-impurity Josephson junction: Variational wave functions 
        and slave-boson mean-field theory}, 
	Phys. Rev. B \textbf{62}, 6687 (2000).
	
\bibitem{fnrg}
	We use open-access Budapest Flexible DM-NRG code, 
	\url{http://www.phy.bme.hu/dmnrg/};
	O. Legeza, C. P. Moca, A. I. T\'{o}th, I. Weymann, G. Zar\'{a}nd,
	{\em Manual for the Flexible DM-NRG code},
	arXiv:0809.3143 (2008, unpublished).
\bibitem{AndersSchiller}	
	F.B. Anders and A. Schiller, 
	{\em Real-Time Dynamics in Quantum-Impurity Systems:
	 A Time-Dependent Numerical Renormalization-Group Approach},
	Phys. Rev. Lett. \textbf{95}, 196801 (2005);
	{\em Spin precession and real-time dynamics in the Kondo model: 
	 Time-dependent numerical renormalization-group study},
	Phys. Rev. B \textbf{74}, 245113 (2006).
\bibitem{dm-nrg}	
	A. Weichselbaum and J. von Delft, 
	{\em Sum-Rule Conserving Spectral Functions from the 
	Numerical Renormalization Group},
	Phys. Rev. Lett. \textbf{99}, 076402 (2007).
\bibitem{Z}
	W.C. Oliveira and L.N. Oliveira, 
	{\em Generalized numerical renormalization-group method to calculate 
        the thermodynamical properties of impurities in metals},
	Phys. Rev. B \textbf{49}, 11986 (1994).

\bibitem{Haldane}
	F.D.M. Haldane,
	{\it Scaling Theory of the Asymmetric Anderson Model},
	Phys. Rev. Lett. \textbf{40}, 416 (1978).

\bibitem{fano1961}
   U.\ Fano, 
   {\em Effects of configuration interaction on intensities and phase shifts}, 
   Phys. Rev. {\bf 124}, 1866 (1961).

\bibitem{miroshnichenko2010}
   A.E.\ Miroshnichenko, S.\ Flach, and Y.S.\ Kivshar, 
   {\em Fano resonances in nanoscale structures}, 
   Rev. Mod. Phys. {\bf 82}, 2257 (2010).

\end{thebibliography}
\end{document}